\documentclass[12pt]{report}

\usepackage[utf8]{inputenc} 
\usepackage{graphicx}
\graphicspath{ {images/} }
\usepackage{caption}
\usepackage{subcaption}
\usepackage{float}
\usepackage[width=150mm,top=35mm,bottom=25mm,bindingoffset=6mm]{geometry}
\usepackage{fancyhdr}
\renewcommand{\headrulewidth}{0pt}
\usepackage[onehalfspacing]{setspace}

\usepackage[backend = biber, style=numeric,sorting=none]{biblatex}
\usepackage{amsmath}
\usepackage{latexsym}
\usepackage{amssymb}
\usepackage{mathtools}  
\usepackage{amsfonts}  
\usepackage{mathrsfs}

\usepackage{hyperref}

\let\oldref\ref
\renewcommand{\ref}[1]{(\oldref{#1})}

\title{Asymptotic Symmetries in Electrodynamics and Kalb-Ramond Theory}
\author{Maya Chaudhuri}
\date{5 July 2022}

\begin{document}
\pagenumbering{roman} 

\begin{titlepage}
    \begin{center}
        
        \LARGE
        Bachelor's Thesis
            
        \vspace{0.5cm}
              
        \rule{\textwidth}{1.5pt}
        \LARGE
        \textbf{Asymptotic Symmetries in Electrodynamics and Kalb-Ramond Theory}
        \rule{\textwidth}{1.5pt}
           
        \vspace{0.5cm}
              
        \large
        Department of Physics \\
        Ludwig-Maximilians-Universität München 
        
        \vfill
        
        \Large
        Maya Chaudhuri
        
        \vfill
        
        \large
        Munich, August 8 \textsuperscript{th}, 2022
              
        \vfill
        
        \includegraphics[width = 0.4\textwidth]{sigillum.png}
        
        \vfill
        
        \large
        Supervised by Prof. Dr. Ivo Sachs
        
    \end{center}
    
\end{titlepage}
\newpage\null\thispagestyle{empty}\newpage
\begin{titlepage}
\begin{center}
    
\LARGE
Bachelorarbeit
    
\vspace{0.5cm}
      
\rule{\textwidth}{1.5pt}
\LARGE
\textbf{Asymptotische Symmetrien in der Elektrodynamik und Kalb-Ramond Theorie}
\rule{\textwidth}{1.5pt}
   
\vspace{0.5cm}
      
\large
Fakultät für Physik \\
Ludwig-Maximilians-Universität München 

\vfill

\Large
Maya Chaudhuri

\vfill

\large
8. August 2022
      
\vfill

\includegraphics[width = 0.4\textwidth]{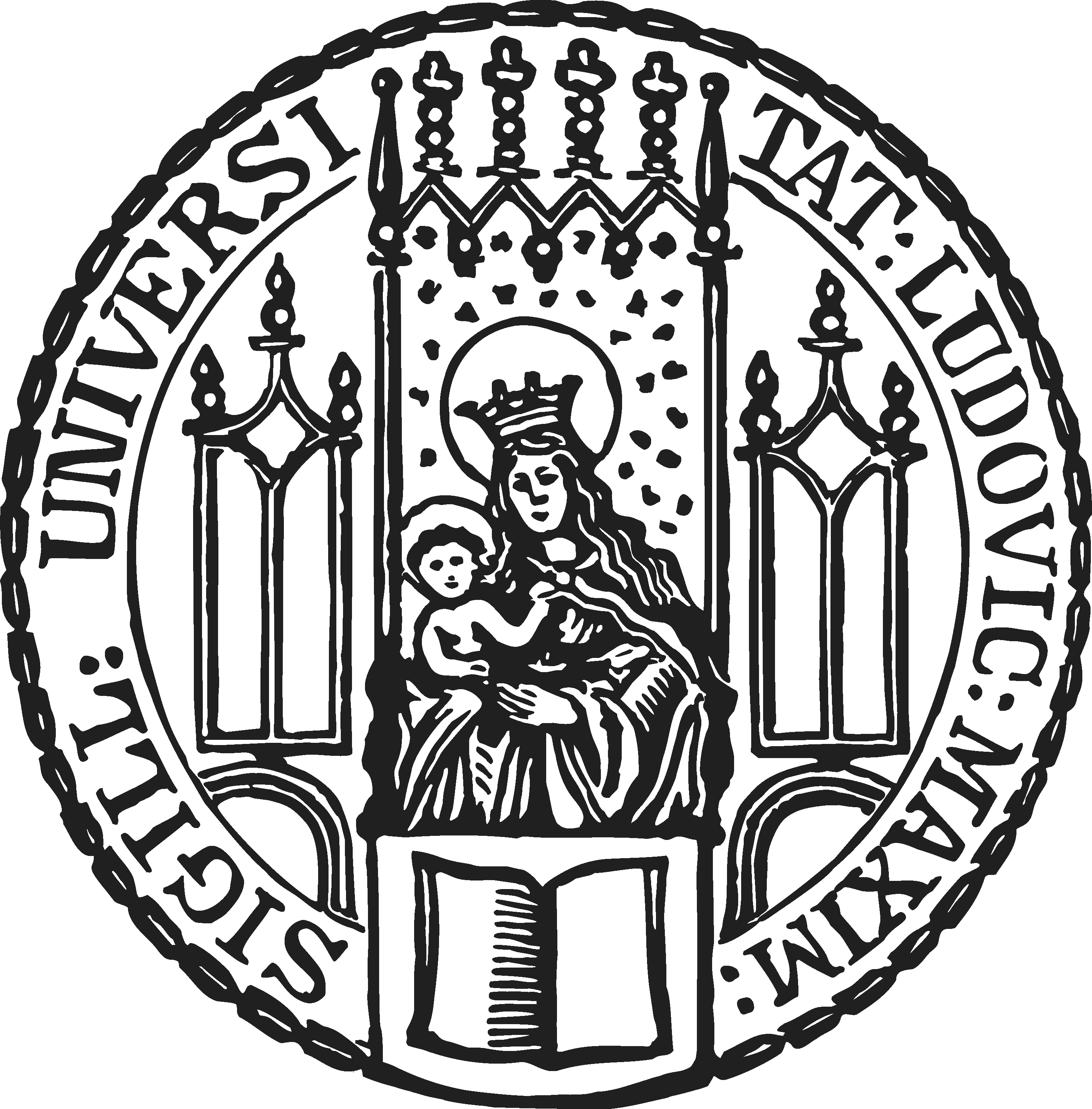}

\vfill

\large
Betreut von Prof. Dr. Ivo Sachs

\end{center}
\end{titlepage}
\newpage\null\thispagestyle{empty}\newpage

\fancyhf{} 
\fancyhead[RO,R]{\thepage} 
\renewcommand{\headrulewidth}{0pt}

\begin{center}
    \textbf{Abstract}
\end{center}
Asymptotic symmetries are residual gauge transformations that preserve boundary conditions and act non-trivially at infinity, i.e. they relate physically distinct states. They are obtained by fixing a gauge,  specifying the large-$r$ fall-off behaviour of the fields and computing the generating Noether charge, which is non-zero.
In this thesis, we aim to find the asymptotic symmetries of the Kalb-Ramond field in four dimensions at future null infinity. The Kalb-Ramond field is a generalization of Maxwell's electrodynamics in which the gauge fields are described by two-forms instead of one-forms.

We start by reviewing the asymptotic symmetries of electrodynamics in four dimensional Minkowski spacetime at future null infinity. After fixing the retarded radial gauge and specifying the fall-off conditions on the gauge fields, we compute the generating charges of the asymptotic symmetries. We use the covariant phase space approach and extend the field space by introducing new independent fields that follow from decomposing the gauge field into a pure gauge part and a gauge-invariant time-dependent part.

In the original contribution of this thesis, we investigate the asymptotic symmetries of the Kalb-Ramond field at future null infinity. We motivate the fall-off conditions by demanding the finiteness of energy, momentum, angular momentum and charge flux through future null infinity. We expand the gauge fields in ``radial" and Lorenz gauge and compute the generating charges. Using the duality between the Kalb-Ramond theory and the scalar field in two dimensions, we again derive the fields' fall-off conditions and compare them to the ones obtained above. 

Our findings can be summarized as follows:

The different gauges yield two similar generating charges, however, the charge obtained in the ``radial" gauge vanishes at infinity. This result might indicate that the fall-off conditions are too strict in this gauge.

Interestingly, we observe a consistency in the asymptotic behaviours of Kalb-Ramond and scalar field theories. Even after we expanded both fields asymptotically, the fall-off conditions for the Kalb-Ramond field obtained by duality considerations are compatible with those derived from the finiteness conditions above. This might also allow us to address the question asked in \cite{Campiglia2018} about which are the missing asymptotic symmetries generated by the soft charges of scalar fields.

\newpage\null\thispagestyle{empty}\newpage

\chapter*{Acknowledgements}
This project would not have been possible without the support of many people.
I would like to thank my supervisor Prof. Dr. Ivo Sachs, as well as Mart\'{i}n Enr\'{i}quez Rojo and Till Heckelbacher, for their regular and extensive discussions, feedback on my work and guidance throughout this project.
And finally, thanks to my parents, my sister Nina and my cat Milchbart.
\newpage\null\thispagestyle{empty}\newpage

\tableofcontents

\chapter*{Notation and Conventions}
\begin{itemize}
    \item Indices $\alpha, \beta, \mu, ...$ : spacetime indices
    \item Indices $A,B,D, ...$ : angular part of spatial indices
    \item $M$: Minkowski spacetime
    \item $g_{\mu\nu}$: metric on Minkowski spacetime
    \item $\gamma_{AB}$: metric on the unit sphere S
    \item $\nabla_\mu$ : covariant derivative on  $M$
    \item $D_A$ : covariant derivative on $S$
    \item $\Box = g^{\mu\nu} \nabla_\mu \nabla_{\nu}$ : d'Alembert operator on $M$
    \item $d\Gamma$: volume form on $S$, e.g. $d\Gamma = sin(\theta) d\theta \wedge d\phi$
    \item $\boldsymbol{\theta}$: angular coordinates, e.g., $\boldsymbol{\theta} = (\theta, \phi)$
    \item $\delta_S (\boldsymbol{\theta}, \boldsymbol{\theta}^\prime)$: Dirac delta on the unit sphere, normalized s.t. $\int_S \delta_S (\boldsymbol{\theta}, \boldsymbol{\theta}^\prime) d\Gamma = 1$
    \item $\Gamma$ : phase space
    \item $\mathfrak{F}(\Gamma)$: space of smooth functions on $\Gamma$
    \item $d$ : exterior derivative on spacetime
    \item $\delta$ : exterior derivative on configuration, solution, and phase space
    \item $\wedge$: wedge product on phase space
    \item $i_X$: interior product with respect to vector field $X$
    \item Big-$\mathcal{O}$-notation: $f(x) \in \mathcal{O} (g(x)) \textrm{ iff } lim_{x\rightarrow \infty} |\frac{f(x)}{g(x)}| < \infty$
    \item Calligraphic letters for leading order terms in $r$: e.g. if $A(r) \in \mathcal{O} (\frac{1}{r})$ : $\mathcal{A} = lim_{r \rightarrow \infty} r A(r)$
    \item Upper indices $\pm$ for limits $u \rightarrow \pm \infty$

\end{itemize}
We will work in units $c= \hbar = 1$

\noindent The metric signature is $(+,-,-,-)$.

\noindent We will use the Einstein summation convention for all kind of numerical indices occuring once as a sub- and once as a superscript index. There is no summation over $\theta, \phi$ implied as they are reserved for the respective components in spherical coordinates.

\noindent We raise and lower all indices using the spacetime metric only.In particular, for angular indices $A^A = g^{AB} A_B \neq \gamma^{AB} A_B$.

\listoffigures
\newpage\null\thispagestyle{empty}\newpage

\chapter{Introduction}
\pagenumbering{arabic} 

Asymptotic symmetries were originally studied in 1962 by Bondi, Metzner, van der Burg and Sachs \cite{Bondi1962, Sachs1962a, Sachs1962b} in the context of general relativity. They were interested in the gravitational field of isolated or weakly radiating objects, like a planet orbiting around a star, at asymptotically flat spacetime.
As one follows the gravitational waves and moves far away from the sources, the metric tensor approaches that of a flat Minkowski spacetime. One then specifies asymptotic fall-off conditions on the radiation at infinity. 
The residual symmetry transformations, which keep the fall-off conditions invariant, were called asymptotic symmetries. 
In the 1960s, they expected the residual symmetries to be the symmetries of flat Minkowski spacetime, namely the Poincaré symmetries. Instead, however, they found the infinite-dimensional BMS group, which contains the Poincaré group as a subgroup.

The concept of asymptotic symmetries was recently extended to other theories, like Maxwell's electrodynamics, which is a gauge theory \cite{Strominger2013}.
The procedure to find the asymptotic symmetries of a gauge theory starts by fixing a gauge. One then specifies the relevant set of solutions to the equations of motion by placing fall-off conditions on the gauge fields, i.e. by choosing how they reduce to a flat configuration as one goes very far from the source. The asymptotic symmetries are the non-trivial transformations that preserve the set of solutions. It turns out that these symmetries often lead to an infinite-dimensional enhancement of the global symmetries, just like in gravity. \cite{Strominger2013, Barnich2013, He2014}

Why are asymptotic symmetries relevant?
Asymptotic symmetries imply observable consequences: memory effects. Their connection is specified in the infrared triangle, which relates asymptotic symmetries, memory effects and soft theorems. Major contributers to this concept were Weinberg (soft theorems) \cite{Weinberg1964, Weinberg1965}, Bieri, Pasterski, Zeldovich, Christodoulou (memory effects) \cite{Bieri2010, Pasterski2015, Zeldovich1974, Christodoulou1991}  and Strominger (asymptotic symmetries) \cite{Strominger2013, Strominger2018}.
They discovered that the charges corresponding to the asymptotic symmetries \footnote{This correspondence is described in Noether's theorem (c.f. chapter 2.1).} are equivalent to a Ward identity \footnote{A Ward identity is the quantum version of classical current conservation associated with a continuous symmetry by Noether's theorem.}, that translates into soft theorems \cite{Zeldovich1974}. 
Furthermore, it has been observed that the passage of radiation near a charge can leave a permanent observable imprint on the physical properties of the charge, which is called the memory effect. Strominger and Zhiboedov \cite{Strominger2013} discovered that the passing radiation changes the underlying gauge field by a vacuum transition, i.e. the passage of radiation induces a transformation from one radiative vacuum to another inequivalent vacuum, both of which are connected by an asymptotic symmetry. Memory effects are, therefore, an observable consequence of the concept of asymptotic symmetries.

Asymptotic symmetries are also relevant in light of the black hole entropy problem \cite{Carlip2018}. The no-hair theorem states that a black hole can be characterized only by its mass, charge and angular momentum \cite{Ruffini1971}. The problem is that these three global charges can not account for the whole entropy of the black hole. One can now introduce asymptotic symmetries at the event horizon of the black hole. The associated infinitely many conserved charges may provide, at least partially, the remaining degrees of freedom to describe the entropy.

This thesis will address the asymptotic symmetries of electrodynamics and the Kalb-Ramond theory at future null infinity in four dimensions.
In chapter \ref{chap2}, we introduce essential concepts needed for later parts of the thesis. We begin with the notion of symmetries in physics in chapter \ref{chap21}. Then, chapter \ref{chap22} gives an overview of the Minkowski spacetime at future null infinity and, for this reason, introduces the conformal compactification of spacetimes. Next, we present the covariant phase space formalism, which is needed to construct a phase space on a null infinity in chapter \ref{chap23}. 

With these concepts, we review the asymptotic symmetries of electrodynamics in chapter \ref{chap3}. After fixing a gauge and imposing reasonable boundary conditions, we formulate the phase space for future null infinity and compute the conserved charges.

Eventually, we perform a similar analysis on the Kalb-Ramond theory in \\ chapter \ref{chap4}. This is the original part of this thesis. The Kalb-Ramond theory is a generalization of electrodynamics, where gauge fields are described by two-forms instead of one-forms in electrodynamics. An analysis of asymptotic symmetries for these fields of higher ranks would be interesting, and while it has been investigated for spatial infinity \cite{Afshar2018}, it has not been done for null infinity, to our knowledge.

In four dimensions, we calculate the associated Noether charges by fixing a gauge and again imposing boundary conditions on the Kalb-Ramond fields. We will also comment on the duality between the Kalb-Ramond theory and the scalar field theory in four dimensions, which will turn out to be very interesting.

Finally, in chapter \ref{chap5}, we will conclude this thesis with a discussion and point out possibilities for future research, appendices and the bibliography.

\chapter{Preliminaries} \label{chap2}

\section{Symmetries in Theoretical Physics} \label{chap21}

A symmetry describes a characteristic of a physical system that is invariant under some symmetry transformation \cite{Feynman2010}. 
A prominent example is the fact that the speed of light remains the same in all reference frames.

In this section we will review global symmetries, local and gauge symmetries, Noether's theorem and asymptotic symmetries. This is the basis for the later analysis.

\subsection{Global, Local and Gauge Symmetries}

    A global symmetry is one that is does not depend on spacetime coordinates. In other words, it keeps a property invariant under a transformation that is applied simultaneously at all points of spacetime.

    In contrast, a local symmetry depends on spacetime coordinates. It therefore keeps a property invariant when a possibly different symmetry transformation is applied at each point of spacetime. Requiring a local symmetry implies the introduction of an additional field, the gauge field, to keep the theory invariant under such transformations. One can see this as the gauge field ``communicating'' phase changes from one point to another. The mathematical expression for this is a connection field. In electrodynamics this connection field is the photon field $A_\mu$ that carries the electromagnetic force.
    
    A better term for ``gauge symmetry" would be ``gauge invariance", because ``gauge symmetries" are technically not symmetries. Rather they are a redundancy in the description of a theory, that can be removed through gauge fixing \cite{Henneaux1986}. 
    Any gauge invariance of the Lagrangian is equivalent to a constraint in the Hamiltonian formalism, i.e. a non-trivial relation among the equations of motion  \cite{Banados2017}.
    A gauge transformation does not relate physically distinct states. Instead it relates two identical solutions, that only differ in their mathematical presentation.
    A gauge theory has more undetermined variables than equations.
    The process of gauge fixing is to select one of the many equivalent solutions related via gauge transformation, by imposing a condition on the gauge field. Fixing a gauge removes the ``unphysical" degrees of freedom, i.e. choices of variables that are irrelevant to the physics.
    
    In contrast, a global symmetry is a ``true" symmetry of the system. It does not reduce the degrees of freedom of the system, but corresponds to conserved quantities, as we will see in the section about Noether's theorem.

\subsection{Noether's Theorem}

    Noether's theorem was introduced in 1918 by Emmy Noether \cite{Noether1971}. We will follow the accessible review of \cite{Banados2017} in this section.
    
    Nother's theorem states that each continuous symmetry of a physical system implies the conservation of some physical quantity of that system. Vice versa, each conserved quantity corresponds to a symmetry.
    
    The theorem can be derived from a relation between {symmetry variations} and {on-shell variations} of an action.
    
    We start with an action of a field theory in four dimensions
    \begin{equation}
        S[\phi] = \int d^4 x L(\phi, \partial_\mu \phi)
    \end{equation} 
    with fields $\phi (x)$ and the Lagrange density $L(\phi, \partial_\mu \phi)$.
    
    The Euler-Lagrange equations are 
    \begin{equation}
      \partial_\mu (\frac{\partial L} {\partial \phi_{, \mu}}) - \frac{\partial L}{\partial \phi} = 0 ,
    \end{equation}
   introducing the notation $\phi_{, \mu} \equiv \partial_\mu \phi$. 
    
    What is a symmetry variation? A symmetry variation is a function $\delta_s \phi$ such that for any $\phi$, the action is invariant up to a boundary term $K^\mu$. The subscript $s$ symbolizes that this is a symmetry transformation.
    The variation of the action is 
    \begin{equation} \label{e1}
        \delta S[\phi, \delta_\epsilon \phi] \equiv S[\phi + \delta_s \phi] - S[\phi] = \int d^4 x \partial_\mu K^\mu .
    \end{equation}
    The notation $\delta S$ denotes the variation of the action under the symmetry.
    
    Given an action $S[\phi]$, all functions $\delta_s \phi$ satisfying equation \ref{e1} for arbitrary $\phi$ are symmetries.
    
    What is an on-shell variation? In contrast to the symmetry variation, in this case, the variations $\delta \phi$ are arbitrary while the fields $\phi$ are constrained by the Euler-Lagrange equations, i.e. they are evaluated on-shell, which is denoted by $\bar{\phi}$. The variation of the action is then
    \begin{equation}
    \begin{split}
         \delta S[\bar\phi, \delta \phi] &= \int d^4x (\frac{\partial {L}}{\partial \phi} \delta \phi + \frac{\partial {L}}{\partial \phi_{,\mu}} \delta \phi_{,\mu}) \\
        &= \int d^4x (\frac{\partial {L}}{\partial {\phi}} - \partial_\mu (\frac{\partial {L}}{\partial \phi_{,\mu}})) \delta \phi + \int d^4x \partial_\mu (\frac{\partial {L}}{\partial \phi_{,\mu}} \delta {\phi}) \\
        &= \int d^4x \partial_\mu (\frac{\partial {L}}{\partial \phi_{,\mu}} \delta {\phi}) .
    \end{split}
    \end{equation}

    One can see that the bulk of the variation vanishes on-shell. What remains is just a total derivative.
    
    Recall, that the symmetry variation was defined as
    \begin{equation} \label{e25}
        \delta S[\phi, \delta_\epsilon \phi] = \int d^4 x \partial_\mu K^\mu ,
    \end{equation}
    which is valid for any $\phi$, in particular for $\bar\phi$ 
    and the on-shell variation as
    
    \begin{equation} \label{e26}
        \delta S[\bar\phi, \delta \phi] = \int d^4x \partial_\mu (\frac{\partial {L}}{\partial \phi_{,\mu}} \delta {\phi}) ,
    \end{equation}
    which is valid for any $\delta \phi$, in particular for $\delta_s \phi$.
    
    Combining the symmetry and on-shell variation leads to Noether's theorem.
    
    By inserting $\bar\phi$ into \ref{e25} and $\delta_s \phi$ into \ref{e26} and substracting the two equations the left sides vanish and we can read off the following equation
    
    \begin{equation}
        \partial_\mu J^\mu = 0 ,
    \end{equation}
    
    where 
    \begin{equation}
        J^\mu \equiv \frac{\partial L} {\partial \phi_{,\mu}} \delta \phi - K^\mu .
    \end{equation}
    
    In other words, given the symmetry transformation $\delta_s \phi$, the quantity $J^\mu$ is conserved. This is exactly Noether's first theorem.
    
    An important example of a conserved quantity corresponding to a continuous symmetry is for example the conservation of energy, which is associated to time invariance.

\subsubsection{Example: Scalar Field}

    In the following, we will derive the Noether current and charge of the complex scalar field to illustrate the concept presented above.
    Given the action
    
    \begin{equation}
        S = \int_M d^4 x \sqrt{-g_M} \ g^{\mu\nu}_M \nabla_\mu \phi^* \nabla_\nu \phi , 
    \end{equation}
    
    where $\phi$ is the complex scalar field and $\nabla_\mu$ the covariant derivative on Minkowski spacetime, we can compute the variation
    \begin{equation}
    \begin{split}
        \delta S &= - \underbrace{\int_M d^4 x \sqrt{-g_M} \ g^{\mu\nu}_M \ (\delta \phi^* \Box \phi + \Box \phi^* \delta \phi)}_{\textrm{bulk term}} \\
        &\mathrel{\phantom{=}} + \underbrace{\int_M d^4 x \sqrt{-g_M} \ g^{\mu\nu}_M \ \nabla^\mu (\delta \phi^* \nabla_\mu \phi + \nabla_\mu \phi^* \delta \phi)}_{\textrm{boundary term}} ,
    \end{split}
    \end{equation}
    with the d'Alembert operator $\Box = \nabla^\mu \nabla_\mu$. From the bulk term one gets the equations of motion $\Box \phi = \Box \phi^* = 0$.
    
    Following the Noether procedure we construct the Noether current.
    The symmetry transformation we want to consider is the global $U(1)$ transformation
    \begin{equation}
    \phi \rightarrow e^{i\epsilon} \phi ,    
    \end{equation}
    
    where $\epsilon$ is a infinitesimal parameter. Using Taylor expansion we can write 
    
    \begin{equation}
        e^{i\epsilon} \phi \approx (1+i\epsilon) \phi = \phi + i\epsilon \phi
    \end{equation}
    
    and identify the transformation
    
    \begin{equation}
        \delta_\epsilon \phi = i \epsilon \phi .
    \end{equation}
    
    This is a symmetry, so the action is invariant under this symmetry transformation: $\delta_\epsilon S = 0$.

    The on-shell variation evaluated at $\bar{\phi}, \bar{\phi}^* \in \mathcal{S}$ is
    
    \begin{equation}
        \delta S [\bar{\phi}, \bar{\phi}^*] = \int_M d^4 x \sqrt{-g}  \ \nabla^\mu (\delta {\phi}^* \nabla_\mu \bar{\phi} + \nabla_\mu \bar{\phi}^* \delta {\phi}) .
    \end{equation}
    
    We now evaluate the symmetry variation on the solution space and the on-shell variation for a symmetry transformation, and demand that the difference of the corresponding expressions must vanish:
    \begin{equation}
        -i\epsilon \int_M d^4 x \sqrt{-g}  \ \nabla^\mu (\delta \bar{\phi}^* \nabla_\mu \bar{\phi} + \nabla_\mu \bar{\phi}^* \delta \bar{\phi}) =0 .
    \end{equation}
    
    Then, the Noether current is defined as
    \begin{equation} \label{scacur}
        j_\mu (\epsilon) = i\epsilon (\nabla_\mu {\phi}^* {\phi} -  {\phi}^* \nabla_\mu {\phi}) .
    \end{equation}
    
    The Noether charge can be obtained by integrating this over a Cauchy surface. Even though our analysis later focuses on future null infinity, we will illustrate our example for the simpler case of a constant time slice $\Sigma_t$ in Cartesian coordinates. 
    The normal vector to $\Sigma_t$ is $(\frac{\partial}{\partial t})^\mu = \delta_t^\mu$ and the directed top form is $d\Sigma^\mu = \delta_t^\mu d^3 x$.
    Then we can write
    
    \begin{equation} \label{scalarNo}
        Q_\epsilon = \int_{\Sigma_t} j_\mu (\epsilon) d\Sigma^\mu = i \epsilon \int_{\Sigma_t} (\nabla_t {\phi}^* {\phi} -  {\phi}^* \nabla_t {\phi}) d^3x .
    \end{equation}

\subsection{Asymptotic Symmetries}

    Noether's theorem states that continuous symmetries lead to conserved currents. In gauge theories such as electromagnetism, the currents associated to gauge transformations vanish, because gauge symmetries are not physical symmetries, but rather redundancies of description \cite{Banados2017}. 
    
    However, if the manifold on which the gauge theory is defined has a boundary and the gauge parameter does not vanish on it, then the associated conserved charge can be non-zero. They are a subgroup of gauge symmetries called improper gauge transformations. Such gauge transformations that do not vanish at infinity are known as asymptotic symmetries and they act non-trivially on the space of states, meaning they relate physically distinct states.
    
    The group of asymptotic symmetries is the set of all symmetry transformations of the field that preserve the asymptotic boundary conditions and possess a non-zero Noether charge \cite{Banados2017}.
    The boundary conditions should be weak enough so that all physically reasonable solutions are allowed, but strong enough so that relevant charges are finite and well-defined. The allowed gauge symmetries are those that respect the boundary conditions. The trivial gauge symmetries are the ones that act trivially on the physical data of the theory. \cite{Strominger2018}
    
    In order to obtain the group of asymptotic symmetries of a gauge theory, one fixes the gauge and specifies the large $r$ fall-off behaviour of the fields. 
    
    Given boundary conditions imposed in a chosen gauge, the asymptotic symmetries are then defined as the residual gauge transformations preserving the boundary conditions and act non-trivially at infinity.

\section{Minkowski Space at Infinity} \label{chap22}

Our analysis will focus on future null infinity. In this chapter we will review the concept of conformal compactification of a spacetime to pave the way for a discussion of the asymptotic behaviour on a manifold.
This section is mainly based on \cite{Schröder2022, Penrose2011, Mitra2017}.

The Minkowski metric in spherical coordinates is given by
\begin{equation} \label{Minmet}
    ds^2 = g_{\mu \nu} dx^{\mu}dx^{\nu} = dt^2 -dr^2 -r^2 (d\theta^2 +sin^2\theta d\varphi^2) = dt^2 -dr^2 -r^2 \gamma_{AB} d\theta^A d\theta^B ,
\end{equation}
where $\boldsymbol{\theta} = (\theta^A, \theta^B)$,  e.g. $\boldsymbol{\theta} = (\theta, \varphi)$,  are the angular coordinates and $\gamma_{AB}$ is the metric on the on the two sphere $S$.
We used the "mostly minus" sign convention for the metric, i.e. the metric signature is $(+,-,-,-)$.

The covariant derivatives are denoted by $\nabla$ on spacetime and $D$ on the two-sphere whose indices can be raised and lowered as
\begin{align}
    \nabla^\nu = g^{\mu\nu} \nabla_\mu , && D^A= \gamma^{AB} D_B.
\end{align}
The coordinates are restricted to $t \in (-\infty, \infty)$, $r \in (0, \infty)$, $\theta \in (0, \pi)$ and \\
$\varphi \in [0,2 \pi)$.

There are five possible infinities one could reach by considering limits of the coordinates:
\begin{itemize}
    \item Spatial infinity ${i}^0$ is reached by taking the limit $r \rightarrow +\infty$ while t stays finite.
    \item Future and past timelike infinity ${i}^+$, ${i}^-$ are reached by taking the limits \\
    $t \rightarrow +\infty$ or $t \rightarrow -\infty$, respectively, while $r$ stays finite.
    \item Future and past null infintiy $\mathcal{I}^+$, $\mathcal{I}^-$ are reached by taking the limits $r \rightarrow +\infty$ or $r \rightarrow -\infty$, respectively, while the retarded time $u := t-r$ or advanced time $v :=t+r$ converge to finite values, respectively.
\end{itemize}

To work more intuitively with those infinities it is helpful to define them as finite loci on a new, unphysical manifold, where they are at a finite distance with respect to the new, unphysical metric.  This concept is called conformal compactification. The main idea is to map the original, physical Minkowski spacetime manifold onto a finite region. This is accomplished by a conformal transformation that diverges at the boundaries. \footnote{A conformal transformation preserves angles.} \cite{Penrose2011}

Even though this transformation does not preserve distances, it does leave the causal structure \footnote{The causal structure tells us whether two points are spacelike, null or timelike separated.}  unaffected and light rays still propagate at 45 degrees.  \cite{Strominger2013}

\begin{figure}
    \centering
    \includegraphics[scale=0.6]{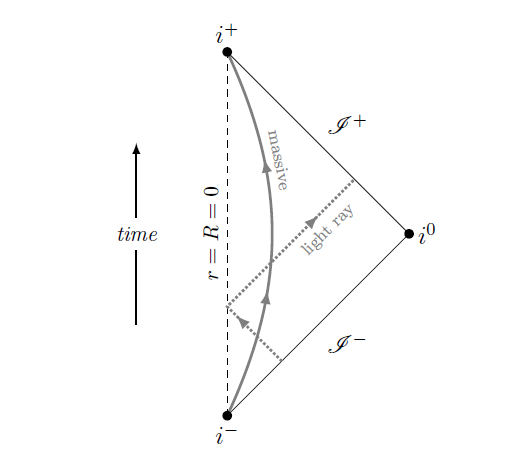}
    \caption[The Penrose diagram for Minkowski spacetime]{The Penrose diagram of Minkowski spacetime, suppressing the angular coordinates and the coordinate singularity $r=0$. The physical spacetime is contained in the interior of a triangle, whose boundaries are timelike, spacelike and null infinities. The figure shows the trajectories of massive particles traveling from $i^-$ to $i^+$. Light rays travel from $\mathcal{I}^-$ to $\mathcal{I}^+$.}
    \label{fig:penrose}
\end{figure}

Let us consider $M$ to be a general, physical spacetime with the metric $g$. Now, we transform the metric conformally $g \rightarrow \tilde{g} = \Omega^2 g$ to condense the entire spacetime $M$ into a finite region. This region can then be represented on a two-dimensional diagram, called the Penrose-Carter diagram (figure \ref{fig:penrose}).
We denote the new, unphysical manifold by $\tilde{M}$ with the metric $\tilde{g}$.

By demanding that $\Omega=0$ on $\partial \tilde{M}$,  the infinity of $M$ can now be expressed as the finite boundary $\partial{\tilde{M}}$ of $\tilde{M}$. \cite{Penrose2011, Mitra2017, Tanzi2017}

Furthermore, we can identify $M$ as the interior of the new, unphysical spacetime $\tilde{M} \backslash \partial \tilde{M}$. 
In other words, infinities are now actual boundaries of a new, unphysical manifold and are at finite distances with respect to the new, unphysical metric. \cite{Penrose2011} 

\begin{figure}
    \centering
    \label{fig1}
    \includegraphics[scale=0.5]{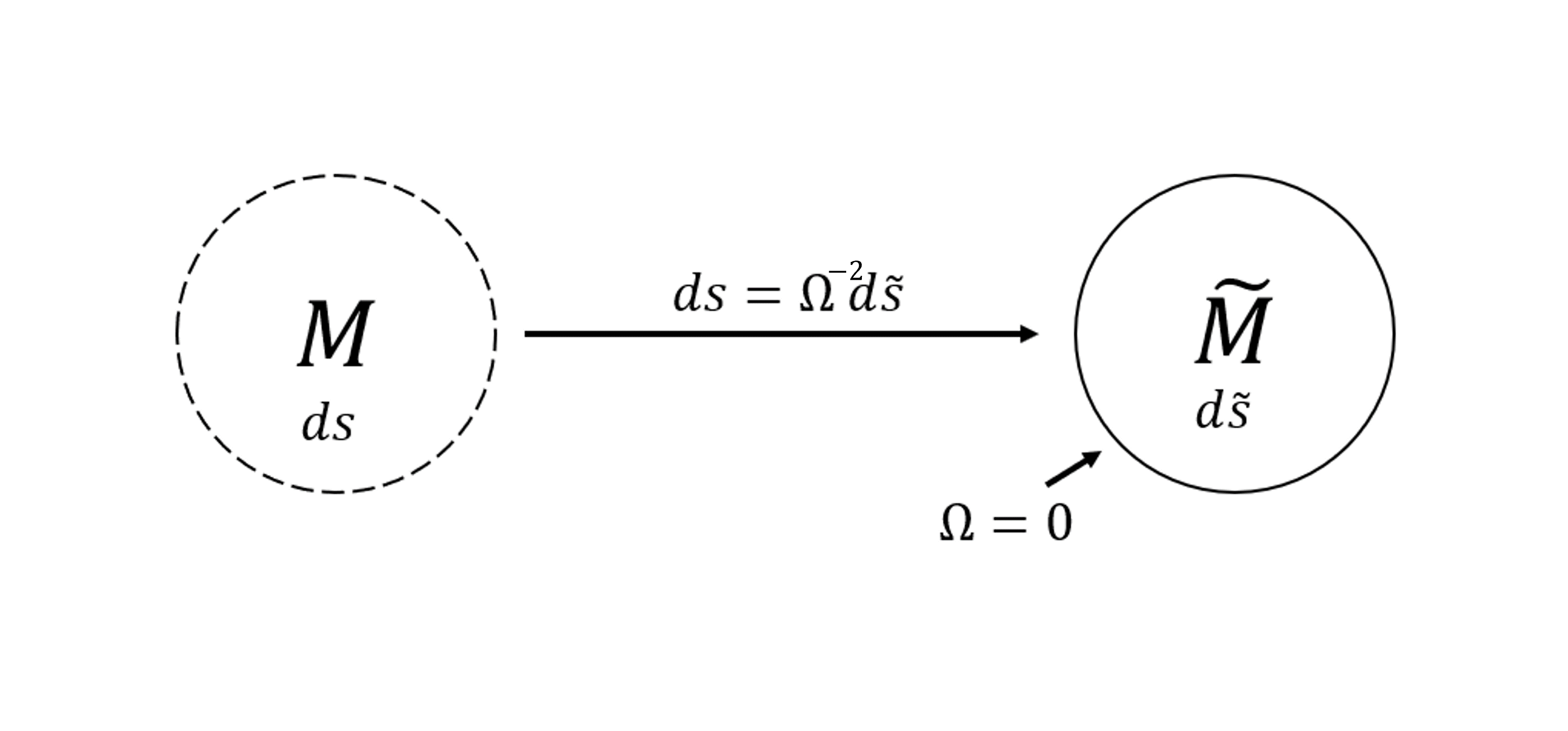}
    \caption[Conformal transformation of a spacetime]{The infinite physical spacetime $M$ is mapped into an unphysical, finite manifold $\tilde M$ by a conformal transformation $\Omega$. The conformal transformation vanishes on the boundary of the new manifold that can be identified with the infinity of $M$. (Illustration based on \cite{Penrose2011})
}
    \label{fig:conformal}
\end{figure}

Now, we want to apply this directly to the Minkowski spacetime, that we will from now on denote by $M$ again. 
To find a conformal transformation, we first define the compact coordinates $T$ and $R$ \cite{Mitra2017}.
We've already introduced the retarded and advanced time $u$ and $v$ , respectively, above. They are null coordinates \footnote{ A null-coordinate $u$ satisfies $g^{-1}(du,    du)=0$ \cite{Zwiebach2009}}.
\begin{align}
    u=t-r \ \textrm{(retarded time)}, && v=t+r \ \textrm{(advanced time)}, && u,v \in \mathbb{R}, u<v .
\end{align}

We now want to bring the infinities of $u$ and $v$ to finite values. To this end, we define the coordinates $U$ and $V$ by
\begin{align}
     U=arctan(u), &&  V=arctan(v), && U,V \in (-\frac{\pi}{2}, \frac{\pi}{2}), U<V .
\end{align}

Now, we can construct the compact coordinates
\begin{align}
    T=V+U, && R=V-U, && (T+R), (T-R) \in (-\pi, \pi), R<0 .
\end{align}

The total coordinate transformations are:
\begin{equation}
\begin{split}
    t(T,R) = \frac{sin(T)}{cos(R) + cos(T)} , \\
    r(T,R) = \frac{sin(R)}{cos(R) + cos(T)} .
\end{split}
\end{equation}

With the conformal transformation $\Omega(T,R) = cos(T) + cos(R)$, the line element becomes
\begin{equation}
    ds^2 = \Omega^{-2} d\tilde{s}^2= \Omega^{-2} (dT^2 - dR^2 - sin^2(R) \gamma_{AB} d\theta^A d\theta^B) .
\end{equation}

The infinities of Minkowski spacetime $M_4$ are now actual boundaries of the new, unphysical manifold and can be represented in the Penrose diagram in figure \ref{fig:penrose}.

\begin{itemize}
    \item Spatial infinity ${i}^0$ is a point at $(T,R)= (0, \pi)$.
    \item Future and past timelike infinity ${i}^+$, ${i}^-$:
    \begin{itemize}
        \item $i^+$ is a point at $(T,R) = (\pi, 0)$ .
        \item $i^-$ is a point at $(T,R) = (-\pi, 0)$.
        \end{itemize}
    \item Future and past null infinity $\mathcal{I}^+$, $\mathcal{I}^-$:
    \begin{itemize}
        \item $\mathcal{I}^+$ is a null hypersurface with the line segment $T(R) = \pi -R, R \in (0, \pi)$.
        \item $\mathcal{I}^-$ is a null hypersurface with the line segment $T(R) = R - \pi , R \in (0, \pi)$. 
    \end{itemize}
\end{itemize}

Every massive particle's path begins ${i}^{-}$ and ends at ${i}^{+}$, whereas every light rays starts at $\mathcal{I}^-$ and ends at  $\mathcal{I}^+$  \cite{Tanzi2017}.
Therefore, $\mathcal{I}^+$ represents the surface where all null trajectories of Minkowski space have their future endpoints.

For the analysis at future null infinity the coordinates $t$ and $r$ are unsuitable, so we choose the retarded Bondi coordinates, which are more useful.
Using retarded Bondi coordinates $(u,r, \boldsymbol{\theta})$ with retarded time $u=t-r$ , the metric \ref{Minmet} can be written as
\begin{equation}
    ds^2 = du^2 + 2dudr - r^2 \gamma_{AB} d\theta^A d\theta^B .
\end{equation} 

Future null infinity is reached by setting the advanced time $v = \infty$. To find the remaining coordinates parametrizing $\mathcal{I}^+$ we rewrite $v=t+r=u+2r$. If we keep $u$ constant, it follows that $r \rightarrow \infty$. Thus, the coordinates describing future null infinity are $(u, \boldsymbol{\theta})$.

Additionally, one could now consider the limit  $u \rightarrow \infty$, leading to the two sphere $\mathcal{I}^+_+$. Analogously, in the limit $u \rightarrow -\infty$ , we arrive at the two sphere $\mathcal{I}^+_-$. 
Note, that they are not the same as ${i}^+$ and ${i}^0$. This is because only massless particles can reach $\mathcal{I}^+$, while only massive particles can reach $i^+$ and $i^0$. A massive particle can therefore never reach $\mathcal{I}^+_+$ or $\mathcal{I}^+_-$. \cite{Schröder2022}

We want to integrate vector fields over $\mathcal{I}^+$. To be able to perform such integrals, we need the directed surface element on future null infinity.
The null normal vector and volume element on $\mathcal{I}^+$ is $n_{\mu}=\delta^{u}_{\mu} + 2 \delta^{r}_{\mu}$  and $n^{\mu}=2\delta_{u}^{\mu} - \delta_{r}^{\mu}$  and 
\begin{equation} \label{sigma}
    d\Sigma^{\mu} = \lim \limits_{r \to \infty} r^2  (\delta^{\mu}_{u} - \frac{1}{2} \delta^{\mu}_{r}) \ du \wedge d\Gamma
\end{equation}  
with $d\Gamma$ as the volume form on the two-sphere $S$ .
A derivation of this can be found in \cite{Schröder2022, RojoSchröder2022}.

\section{The Covariant Phase Space Formalism} \label{chap23}

In this section, we introduce the relevant aspects of the covariant phase space formalism, which are required later in order to construct a phase space for the fields on null infinity. 

A phase space describes all posible states of a system. Each state represents one unique point in the phase space.
In classical mechanics, the state contains not only the position $q^i$ of each object in the system (this would be the configuration space), but also their momenta $p_i$. Thus the phase space can be parametrized by the local coordinates $(q^i, p_i)$, which are the the values of position and momentum at a given time $t$. The Hamiltonian and corresponding Poisson brackets are also defined on constant time-slices.

The problem is that this description refers to a fixed point in time and is therefore not covariant, which can be a problem. Our analysis takes place at future null infinity, which means we want to work on Cauchy surfaces that are not constant-time slices.

To solve this problem we need a way to describe the dynamics of a system while preserving covariance and being independent of constant time slices.
The method to achieve it is the "covariant phase space formalism".
The dynamics of a system is typically described in terms of a Lagrangian, and the covariant phase space formalism is a "recipe" that allows us to construct the phase space of a theory given the Lagrangian. 
Within this formalism, one can define a phase space and Poisson brackets for arbitrary Cauchy surfaces. \cite{He2020}

The main idea is to consider the phase space to be the space of solutions of the equations of motion to a given theory instead of the space of initial conditions, in order to be independent of a particular time slice. As we will see, there can arise some problems caused by infinite dimensional phase spaces and gauge invariance.

The covariant phase space formalism was introduced by Wald, Lee and Iyer \cite{Wald2000, Lee1990, Iyer1994}, based on earlier works by Crnkovic and Witten \cite{Crnkovi1986}.
A review on this formalism can be found in \cite{Schröder2022, He2020} and \cite{Gieres2021} , which we base ourselves on.

\subsection{Phase Space, Symplectic form, Hamiltonian Vector Fields and Poisson bracket}

In this section, we introduce the symplectic form and Hamiltonian vector fields in order to define the Poisson brackets.

A phase space is a symplectic manifold $\Gamma$. A symplectic manifold $\Gamma$ is a manifold equipped with a symplectic form $\Omega$. A symplectic form is a closed and non-degenerate two form: \cite{He2020}
\begin{eqnarray}
    \textrm{Closed: } && \delta \Omega = 0 . \\
    \textrm{Non-degenerate: } && \forall X \in T\Gamma: \ i_X \Omega = 0 \Rightarrow X=0 .
\end{eqnarray}
We denote the tangent bundle over the phase space $\Gamma$ by $T\Gamma$ and its dual bundle by $T^*\Gamma$ . The exterior derivative on the phase spcae $\Gamma$ (c.f. appendix \ref{appendixforms}) is denoted by $\delta$ to not confuse it with the exterior derivative $d$ on the spacetime, which we will use later. The interior product $i_X$ is specified in appendix \ref{appendixforms}.

As a two-form, it takes two vectors as input and maps them to a real number:
\begin{equation}
    \Omega: T\Gamma \times T\Gamma \rightarrow \mathbb{R} .
\end{equation}

We introduce the map  
\begin{equation}
    \hat{\Omega}: T\Gamma \rightarrow T^*\Gamma, X \mapsto -i_X \Omega .
\end{equation}
One can think of this function as "plugging a vector into the first slot of the symplectic form".
The map $\hat{\Omega}$ is injective if and only if the symplectic form $\Omega$ is non-degenerate and bilinear.

\subsubsection{Finite Dimensional Phase Space}

In a {finite-dimensional phase space} it automatically follows that $\hat{\Omega}$ is invertible, because here injectivity automatically implies bijectivity. 
The inverse function $\hat{\Omega}^{-1}: T^*\Gamma \rightarrow T\Gamma$ is then given by
\begin{gather}
    \forall X \in T\Gamma: \hat{\Omega}^{-1} (\hat{\Omega}(X)) = \hat{\Omega}^{-1} (-i_X \Omega) = X , \\
    \forall \chi \in T^*\Gamma : \hat{\Omega} (\hat{\Omega}^{-1} (\chi))=i_{\hat{\Omega}^{-1} (\chi)} \Omega = -\chi .
\end{gather}

Using this inverse function we can introduce
\begin{gather}
    \Omega^{-1}: T^*\Gamma \times T^*\Gamma \rightarrow \mathfrak{F} (\Gamma) , \\
    (\chi, \psi) \mapsto \Omega^{-1} (\chi, \psi) = \chi (\hat{\Omega}^{-1}(\psi)) , \\
    \forall X,Y \in T\Gamma: \Omega^{-1} (i_X \Omega, i_Y \Omega) = -\Omega(X,Y) ,
\end{gather}
 
where $\mathfrak{F}(\Gamma)$ is the space of smooth functions on $\Gamma$. $\Omega^{-1}$ is antisymmetric and bilinear.

Next, we define the Hamiltonian vector field.
A Hamiltonian $H \in \mathfrak{F}(\Gamma)$ maps each point in the phase space to its energy. We can now construct the Hamiltonian vector field $X_H$ on $\Gamma$ via
\begin{equation}
    X_H(f) = \Omega^{-1} (\delta f, \delta H)
\end{equation}
with the arbitrary function or $f \in \mathfrak{F}(\Gamma)$ \cite{Harlow2019}.
This is equivalent to the expression 

\begin{equation}
    i_{X_{H}} \Omega = -\delta H.
\end{equation}

The Hamiltonian vector field can be defined for any two functions $f, g \in \mathfrak{F}(\Gamma)$, such that we can define the poisson brackets between these two arbitrary functions \cite{Harlow2019}:
\begin{equation} \label{pbrack}
    \{f, g\} = -\Omega(X_{f}, X_{g}).
\end{equation}

Equipped with this definition we could for example express the time evolution of a function $f \in \mathfrak{F}(\Gamma)$ via
\begin{equation} \label{timeev}
    \frac{df}{dt} = \{f, H\} = -\Omega(X_f, X_H) = -X_f(H) .
\end{equation}

Let us illustrate this with an example. 
Usually, the phase space is understood as the set of all possible initial conditions of a system at a fixed time. If we look at a free particle, it is sufficient to consider position and momentum $(q,p)$ of the particle. We can use those to parametrize the phase space $\Gamma$. 

The symplectic form then takes the form of $\Omega = \delta p \wedge \delta q$.
The Hamiltonian vector field are then denoted by $X_q = -\frac{\delta}{\delta p}$ and $X_p = \frac{\delta}{\delta q}$.
Using \ref{pbrack} we find the expected Poisson brackets $\{q,q\} = 0$, $\{p,q\} = 1$, $\{p,p\} = 0$.

With \ref{timeev} we end up with the Hamilton equations:  $\frac{\delta q}{\delta t} = \frac{\delta H}{\delta p}$ and $\frac{\delta p}{\delta t} = -\frac{\delta H}{\delta q}$.

\subsubsection{Infinite Dimensional Phase Space}

In order to define the Poisson brackets the non-degeneracy of the symplectic form is especially important. It ensures the existence of $\Omega^{-1}$, which we used to define the Hamiltonian vector fields $X_f, X_g$.

Unfortunately, this is only true in finite dimensional phase spaces. In our case, we are dealing with field theories where we consider infinite dimensinonal spaces.

In {infinite dimensional phase spaces} the condition that $\Omega$ is non-degenerate is not sufficient to define Poisson brackets, because injectivity does not imply surjectivity of a map \cite{Lee1990}. Therefore, we cannot just assume the existence of $\Omega^{-1}$. 
If $\Omega^{-1}$ doesn't exist, it would mean there isn't an associated Hamiltonian vector field $X_f$ for every $f$ \cite{Schröder2022}.

The solution to this problem is to extend the phase space (originally done for electromagnetism in \cite{He2014}), which we review in chapter \ref{chap3}. Until then we will just assume that $\Omega^{-1}$ does exist.

\subsection{Construction of the Covariant Phase Space from the Lagrangian}

How do we find the symplectic form $\Omega$? It can be constructed from the Lagrangian of the theory.

The configuration space $C$ of a field theory on a spacetime $M$ with the fields $\phi^i$ consists of all field configurations that satisfy the boundary conditions of the theory.
The dynamics of the system is described by the Lagrangian form $L$, which is the usual Lagrangian density times the volume form on the spacetime.

Taking the variation of the action $\delta S = \int_M \delta L$ and integrating by parts allows us to write the Lagrangian form as follows \cite{Harlow2019}
\begin{equation} \label{e240}
    \delta L = E_i \delta \phi^i + d\Theta ,
\end{equation}

where $\delta \phi^i$ are variations of the fields on $C$. $\Theta$ is called the symplectic potential and from demanding that the variation of the action vanishes up to a boundary term, we get the equations of motion $E_i(\{\bar{\phi}^j \}) = 0$. \cite{Harlow2019}

Now we want to describe our theory covariantly. To this end, we introduce the solution space $\mathcal{S}$, which is defined as the space of fields $\bar{\phi}^j$ , which satisfy the equations of motion. This is description is covariant. In contrast, the phase space $\Gamma$ was interpreted as the set of all possible different initial conditions at a fixed time, which is a non-covariant otion.
If the initial value problem is well-defined, there is a bijection between $\Gamma$ and $\mathcal{S}$. Every initial value gets mapped uniquely to a solution. 

How do we find the symplectic form on $\mathcal{S}$?
We read off the symplectic potential $\Theta$ from \ref{e240} and define the the so-called symplectic current:

\begin{equation} \label{omega}
    \omega = \delta \Theta ,
\end{equation}
such that $\omega$ is closed.

By integrating $\omega$ over a Cauchy slice $\Sigma \subset M$ \cite{Lee1990} we obtain the presymplectic potential:

\begin{equation} \label{presym}
    \tilde{\Omega}_{\Sigma} = \int_{\Sigma} \omega .
\end{equation}

From \ref{e240} we see that the symplectic potential is only defined up to transformations $\Theta \rightarrow \Theta + dY$. Therefore, the symplectic current is also only defined up to replacements $\omega \rightarrow \omega + \delta dY$. This applies to the presymplectic potential as well: $\tilde{\Omega}_{\Sigma} \rightarrow \tilde{\Omega}_{\Sigma} + \int_{\partial \Sigma} \delta Y$. \cite{Iyer1994} 

For $\tilde{\Omega}_{\Sigma}$ to be a true symplectic form, it must fulfill the non-degeneracy condition. This becomes a problem in gauge theories. Because of the gauge freedom, the initial value problem is not well-defined. Given specified intial values, there are still many different solutions corresponding to different gauges. But without a well-defined initial value problem, we do not have a bijection between $\Gamma$ and $\mathcal{S}$ and therefore can not identify them with each other.

This problem is solved to continue with the construction of the covariant phase space.
A solution to this is to combine all states that are related by a gauge transformation into equivalence classes. The space of the equivalence classes can be identified with $\Gamma$ via a bijection. Fixing a gauge has the same effect, as it maps the equivalence classes to their representatives, which satisfy the gauge condition. Gauge fixing and the subsequent restriction of $\tilde{\Omega}_{\Sigma}$ to $\Gamma$  ensures the non-degeneracy of the presymplectic form.

Note, that this didn't solve the problem of the non-invertiblility of the presymplectic form in infinite dimensional phase spaces. We will cover this in chapter \ref{chap3}.

\subsubsection{Example: Scalar Field}

We return to the example of the complex scalar field from chapter \ref{chap21} to illustrate the covariant phase space formalism.
After computing the conserved Noether charges \ref{scalarNo}, we want to verify that they really are generators of the $U(1)$-symmetry, i.e. we want to find $\{\phi (x), Q_\epsilon\} = \delta_\epsilon \phi (x)$,  $\{ \phi^*(x), Q_\epsilon \} = \delta_\epsilon \phi^* (x)$. 	\footnote{This condition can be derived by considering the quantized theory. The fields are now operators, in particular the quantum field $\hat{\Psi}$ and the generating charge operator $\hat{Q}_\epsilon$. A $U(1)$-transformation can be expressed as,
	$$\hat{\Psi} \rightarrow e^{i\hat{Q}_\epsilon} \hat{\Psi} e^{-i\hat{Q}_\epsilon} \approx \hat{\Psi} - i [\hat{\Psi}, \hat{Q}_\epsilon] = \hat{\Psi} + \delta_\epsilon \hat{\Psi}$$
	where we used the Taylor expansion and denoted the commutator by $[\cdot, \cdot]$. The relation between Poisson brackets in classical systems and commutators in quantized systems is $\{\cdot, \cdot\} \rightarrow i [\cdot, \cdot]$. In the classical formalism, we can therefore read off $\{\Psi, Q_\epsilon\} = \delta_\epsilon \Psi$ \cite{Nair2005}.}

In order to define the Poisson brackets, we apply the covariant phase space formalism.
We recall that the variation of the action was given by 
\begin{equation}
\begin{split}
    \delta S &= - \int_M d^4 x \sqrt{-g} \ g^{\mu\nu} \ (\delta \phi^* \Box \phi + \Box \phi^* \delta \phi) \\
    &\mathrel{\phantom{=}} + \int_M d^4 x \sqrt{-g} \ g^{\mu\nu} \ \nabla^\mu (\delta \phi^* \nabla_\mu \phi + \nabla_\mu \phi^* \delta \phi) .
\end{split}
\end{equation}
From the boundary term we identify the symplectic potential recalling \ref{e240}
\begin{equation}
    \Theta_\mu = \nabla_\mu \phi \delta \phi^* + \nabla_\mu \phi^* \delta \phi .
\end{equation}
Using \ref{omega} we obtain the symplectic current
\begin{equation}
     \omega_\mu = \delta \Theta_\mu = \nabla_\mu \delta \phi \wedge \delta \phi^* + \nabla_\mu \delta \phi^* \wedge \delta \phi .
\end{equation}
With \ref{presym} we compute the presymplectic form
\begin{equation}
    \begin{split}
        \tilde{\Omega}_t &= \int_{\Sigma_t} \omega_\mu d\Sigma^\mu = \int _{\Sigma_t} (\nabla_t \delta \phi \wedge \delta \phi^* + \nabla_t \delta \phi^* \wedge \delta \phi) d^3 x  \\
        &= \int _{\Sigma_t} (\delta \dot{\phi} \wedge \delta \phi^* +  \delta \dot{\phi}^* \wedge \delta \phi) d^3 x ,
    \end{split}
\end{equation}
which is non-degenerate, because we don't have a gauge symmetry in our theory. As stated above, we assume now that the symplectic form is invertible and can therefore identify $\tilde{\Omega}_t = \Omega_t$.

Equipped with a symplectic form, we can start defining the Poisson brackets.
First, we want to find $\{\dot{\phi}^*({x}), \phi({y})\}$.

Let us define the two functions as
\begin{gather}
    f=\int f(x) \dot{\phi}^* (x) d^3x , \\
g=\int g(x) \phi(x) d^3x .
\end{gather}

and the corresponding Hamiltonian vector fields $i_{X_f} \Omega_t = -\delta f$, $i_{X_g} \Omega_t = -\delta g$.
By definition \ref{pbrack} the Poisson bracket is 
\begin{equation}
    \{f,g\} = -\Omega_t (X_f, X_g) .
\end{equation}

We choose the ansatz 
\begin{equation}
    X_f =\int_{\Sigma_t} x_f(\tilde{x}) \frac{\delta}{\delta \phi (\tilde{x})} d^3 \tilde x 
\end{equation}
for the Hamiltonian vector field associated to $f$, where $x_f (x)$ is an arbitrary function.

Recall that it has to satisfy 
\begin{equation}
i_{X_f} \Omega_t = -\delta f  .
\end{equation}

Thus, we demand
\begin{equation}
    i_{X_f} \Omega_t = - \int_{\Sigma_t} x_f (x) \delta \dot{\phi}^* (x) d^3 x \stackrel{!}{=} - \delta f = - \int_{\Sigma_t} f(x) \delta \dot{\phi}^* (x) d^3 x .
\end{equation}

We see that this is solved by $f(x) = \delta^3 (x-y) = x_f (x)$. We obtain $i_{X_f} \Omega = - \delta \dot{\phi}^* (y)$

For $g$ we find $g(x) = \delta^3 (x-z)$ and
 $X_g = - \frac{\delta}{\delta \dot{\phi}^* (z)}$.

 Now we can compute the Poisson bracket
\begin{equation}
    \{ f, g\} =  \{\dot{\phi}^* (y), \phi (z) \} = -i_{X_f} \Omega_t (X_g) = - \delta^3 (y-z) .
\end{equation}

Analogously, we find $\{\dot{\phi} (y), \phi^* (z) \} =  - \delta^3 (y-z)$ . All other Poisson brackets vanish. \cite{Schröder2022}

Finally, we compute the Poisson brackets with the transformation generators to check if they are symmetry generators
    \begin{gather}
        \{ \phi(x), Q_\epsilon \} = - i \epsilon \int_{\Sigma_t} \{ \dot{\phi}^*(\tilde{x}), \phi (x) \} \phi (\tilde{x}) d^3 \tilde{x} = i \epsilon \phi (x) = \delta_\epsilon \phi (x) , \\
         \{ \phi^*(x), Q_\epsilon \} = i \epsilon \int_{\Sigma_t} \phi^* (\tilde{x}) \{ \dot{\phi}(\tilde{x}), \phi^* (x) \}  d^3 \tilde{x} = - i \epsilon \phi^* (x) = \delta_\epsilon \phi^* (x) .
    \end{gather}
    
This is exactly what we postulated. The charges are the generators of the symmetry transformations. Additionally, we see that $\{Q_\epsilon, Q_{\epsilon^\prime}\} = 0$, what is to be expected from $U(1)$- symmetries.

\chapter{Asymptotic Symmetries of Electrodynamics}
 \label{chap3}

In this section, we will analyze the asymptotic symmetries of Maxwell's classical electrodynamics in four dimensional Minkowski space at future null infinity.
This analysis has been performed several times in literature, for example in \cite{Strominger2013, He2014,  Strominger2018, Schröder2022, Mitra2017}.

First, we consider the action of the theory and derive the equations of motion in retarded Bondi coordinates. Next, we fix the radial retarded gauge and specify the large-$r$ behaviour of the field strenghts and gauge fields at future null infinity. We will then construct a phase space at future null infinity using the covariant phase space formalism (c.f. chapter \ref{chap23}). This enables us to define Poisson brackets. Using Noether's theorem we compute the associated charges to the asymptotic symmetries and verify the result with the Poisson brackets.

This section is mostly a summary of chapter 4 of \cite{Schröder2022}.

\section{Action, Equations of Motion and Gauge Fixing}

Let us start by writing down the action for Maxwell's electrodynamics:
\begin{equation}
    S_{ED} = - \int_{M} d^4x \sqrt{-g} (\frac{1}{4} F_{\mu\nu} F^{\mu\nu} + A_{\mu} J^{\mu}) .
\end{equation}

$A_\mu$ is the $U(1)$-gauge field and is coupled to a non-dynamical, conserved and gauge invariant invariant current $J^{\mu}$. $F_{\mu\nu} =\partial_{\mu} A_{\nu} - \partial_{\nu} A_{\mu}$ denotes the field strength, which is invariant under the gauge transformation

\begin{equation}
    A_{\mu} \rightarrow A_{\mu} + \partial_{\mu} \lambda ,
\end{equation} 
where $\lambda$ is a sufficiently differentiable scalar function.

With the conservation of the current $\nabla_{\mu} J^{\mu}$ one can see that the action is invariant under these gauge transformations, which proves that they are symmetries of the action. $\nabla_\mu$ denotes the covariant derivative on $M$.

The variation of the action is
\begin{equation} \label{edactionwo}
    \delta S_{ED} = \underbrace{\int_{M} d^4x \sqrt{-g} (\nabla^{\mu} F_{\mu\nu}- J_{\nu})  \delta A^{\nu}}_{bulk \ term} - \underbrace{\int_{M} \sqrt{-g} \nabla^{\mu} (F_{\mu\nu} \delta A^{\nu})}_{boundary \ term} ,
\end{equation} 

and by demanding the bulk term to be zero, we find the equations of motion
\begin{equation} \label{edeom}
    \nabla^{\mu} F_{\mu\nu} = J_{\nu} .
\end{equation}

Usually, a boundary term can be neglected provided the fields fall off sufficiently fast. However, in gauge theories like electrodynamics, the assumption that fields decay sufficiently fast at infinity is not justified anymore.\footnote{Usually, one uses arguments related to finiteness of energy, momentum and charges to justify that $A\rightarrow 0$ at infinity. These arguments remain unchanged in the presence of a gauge invariance, but we now also have to allow the configurations $A \rightarrow \partial_\mu \lambda$. Consider the term $\int A_\mu J^\mu$ from the action that describes the gauge field coupling to the conserved current. If $A$ is pure gauge, then $\int \partial_\mu \lambda J^\mu = \int \partial_\mu (\lambda J^\mu)$. The current falls off as $r^{-2}$ but the integration measure grows as $r^2$. If $\lambda$ is non-vanishing at infinity, i.e. the gauge field is non-vanishing at infinity, this boundary term is non-zero and could lead to observable differences in the theory. } Therefore, one needs to consider the boundary term. 

Because our analysis takes place at future null infinity, it is useful to express the equations of motion in retarded Bondi coordinates (c.f. chapter \ref{chap22}):
\begin{eqnarray}
    J_u = \nabla^\mu F_{\mu u} = (\partial_u - \partial_r - \frac{2}{r}) F_{ru} - \frac{1}{r^2} \gamma^{AB} D_A F_{Bu} , \\
    J_r= \nabla^\mu F_{\mu r} = (\partial_r + \frac{2}{r}) F_{ur} - \frac{1}{r^2} \gamma^{AB} D_A F_{Br} ,\\
    J_C= \nabla^\mu F_{\mu C} = \partial_r F_{uC} + (\partial_u - \partial_r) F_{rC} - \frac{1}{r^2} \gamma^{AB} D_A F_{BC} .
\end{eqnarray}

Next, we choose a gauge condition to partially gauge fix the theory. 
We implement the radial gauge $A_r = 0$. This implies that  $A_{r} + \partial_{r} \lambda = 0$. Then, the gauge parameter is $\lambda(u,r,\boldsymbol{\theta}) = - \int^{r} A_{r}(u, r^\prime, \boldsymbol{\theta}) dr^\prime$. 
If we plug this into the equations of motion in retarded Bondi coordinates we get
\begin{align}
    J_u = \frac{1}{r^2} ((\partial_u - \partial_r)(r^2 \partial_r A_u) + \partial_u \gamma^{AB} D_A A_B - D^2 A_u) , \\
    J_r = \frac{1}{r^2} \partial_r (- r^2 \partial_r A_u + \gamma^{AB} D_A A_B) ,\\
    J_C = \partial_r (2\partial_u - \partial_r) A_C - \partial_r \partial_C A_u - \frac{1}{r^2} \gamma^{AB} D_A (D_B A_C - D_C A_B) .
\end{align}

We can further fix the residual gauge transformations by employing the retarded gauge $A_u |_{\mathcal{I}^+} = 0$, which leads to the gauge parameter $\lambda (u, \boldsymbol{\theta}) = - \int^{u} A_u |_{\mathcal{I}^+} (u, \boldsymbol{\theta}) du$.

\section{Radial Fall-off Conditions}
To find the asymptotic symmetries, we need to specify the large-$r$ fall-off behavior of the fields. A reasonable condition in flat Minkowski space is to demand that the the energy, momentum and angular momentum fluxes associated to the electomagnetic field are finite through $\mathcal{I}^+$.

This condition can be formalized by defining a corresponding conserved current $\mathscr{J}_\nu = T_{\mu\nu} X^\mu$, where $T_{\mu\nu}$ is the energy-momentum tensor and $X^\mu$ the Killing vector field \footnote{Killing vector fields are infinitesimal generators of isometries. Their flows generate symmetries, i.e. moving points the same distance along a Killing vector will not change their distances. In Minkowski spacetime, the isometry group is the ten dimensional Poincaré group, which contains the Lorentz group (spatial rotations, Lorentz boosts) and the spacetime translations.}. \cite{Schröder2022, Wald2000}

$\mathscr{Q}= \int_{\mathcal{I}^+} \mathscr{J}_\nu d\Sigma^\nu$ is then a conserved quantity and can be interpreted as the flux of this quantity through $\mathcal{I}^+$. \cite{Wald2000}
The condition can be therefore expressed as 
\begin{equation} \label{Killingcur}
    |\int_{\mathcal{I}^+} T_{\mu\nu} X^{\mu} d\Sigma^{\nu} | < \infty .
\end{equation}

The explicit form of the Killing vector field can be found in \cite{Schröder2022, Carroll2004, Wald1984}. 
As an example, we can choose a solution of the Killing vector fields, so that only $X^r$ is non-zero \cite{Schröder2022}. Then the condition takes the form 
\begin{equation} \label{finitness}
    |\int_{\mathcal{I}^+} r^2 (T_{ur} - \frac{1}{2} T_{rr}) |^2 < \infty ,
\end{equation}
where we used the directed surface element on future null infinity \ref{sigma} \cite{Schröder2022}.
From the electromagnetic energy-momentum tensor $T_{\mu\nu} =  -F_{\mu \alpha} F_\nu^{\alpha} + \frac{1}{4} g_{\mu\nu} F_{\alpha\beta} F^{\alpha\beta}$ we obtain explicit expressions of $T_{ur}$ and $T_{rr}$ and we find the radial fall-off $F_{uA} \in \mathcal{O} (r^0)$ \cite{He2020}. 

We can also extract a further condition: $\lim_{u \rightarrow \pm \infty} \mathcal{F}_{uA} = 0$ where we used the notation $\mathcal{F}_{uA} (u, \boldsymbol{\theta}) = \lim_{r \rightarrow \pm \infty} F_{uA} (u,r,\boldsymbol{\theta})$ . In other words, at asymptotically early or late retarded times $u$, no radiation reaches null infinity, which is in accordance with the finiteness of the energy through future null infinity. \cite{Schröder2022}

The boundary conditions for the remaining field strengths can be determined by examining the condition for the finiteness of the charge flux through $\mathcal{I}^+$:
\begin{equation}
    | \int_{\mathcal{I}^+} J_{\mu} d\Sigma^{\mu} | =  | \int_{\mathcal{I}^+} (\nabla^{\nu} F_{\nu \mu}) d\Sigma^{\mu} |   < \infty ,
\end{equation}
where we used the equations of motion \ref{edeom}.

This leads to \cite{Schröder2022,Mitra2017, Strominger2018}
\begin{align}
    F_{ur} =\mathcal{O} (r^{-2}) , && F_{uA} =\mathcal{O} (r^0) , && F_{rA} =\mathcal{O} (r^{-2}) , && F_{AB} =\mathcal{O} (r^0) .
\end{align}

Our analysis doesn't include magnetic charges, i.e. $\nabla \cdot \boldsymbol{B} = 0$, meaning that the Coulombic part of the radial electric field $F_{AB}$ has to vanish at large distances \footnote{In spherical coordinates, we choose the orthonormal frame $\{ \hat{\boldsymbol{r}}, \partial_{\theta} \hat{\boldsymbol{r}}, \partial_{\phi} \hat{\boldsymbol{r}} \}$, where $\hat{\boldsymbol{r}}$ points radially outwards and the two remaining vectors are tangential to the sphere. Then, $E_r = \hat{\boldsymbol{r}} \cdot \boldsymbol{E}, B_r = \hat{\boldsymbol{r}} \cdot \boldsymbol{B}, E_\perp \sim (\partial_A \hat{\boldsymbol{r}}) \cdot \boldsymbol{E}, B_\perp \sim (\partial_A \hat{\boldsymbol{r}}) \cdot \boldsymbol{B}$. With $F_{ti} = E_i, F_{ij} = -\epsilon_{ijk} B_k$ it follows that $F_{ur} \sim E_r, F_{uA} \sim r E_r, F_{AB} \sim r^2 B_r, F_{rA} \sim r(xE_\perp + y B_\perp)$, with $x,y$ being $r$ independent. Therefore, the Coulombic part is $F_{AB}$.} and we therefore determine \cite{Schröder2022}
\begin{equation} \label{magch}
    \mathcal{F}_{AB} = \lim_{r \rightarrow \pm \infty} F_{AB} (u,r, \boldsymbol{\theta}) = 0 . 
\end{equation}

The next step is to find the fall-off conditions for the gauge fields.
We already fixed the retarded raidal gauge $A_r = 0$, $A_u |_{\mathcal{I}^+} = 0$
.
To find the radial fall-off conditions for the gauge fields, we asymptotically expand the fields
\begin{equation}
    A_{\mu} (u,r,\boldsymbol{\theta}) = \sum_{n=0}^{\infty} \frac{A_{\mu}^{(-n)} (u, \boldsymbol{\theta})} {r^n} .
\end{equation}

From the retarded gauge condition $A_u |_{\mathcal{I}^+} = 0, A_r = 0$, we can immediately conclude $A_u \in \mathcal{O} (r^{-1})$.
Using this we can examine the field strength $F_{uA}$ and take its radial limit $\lim_{r \rightarrow \pm \infty} F_{uA} =  \mathcal{F}_{uA} =\lim_{r \rightarrow \pm \infty} (\partial_u A_A - \partial_A A_u)  = \partial_u \mathcal{A}_A$ .
Together with the condition $\lim_{u \rightarrow \pm \infty} \mathcal{F}_{uA} = 0$ from above, it follows that $\mathcal{A}_A$ is $u$-independent.

Moreover, the condition $\mathcal{F}_{AB}= 0$ \ref{magch} implies that $\mathcal{A}_B$ must be of the form $\mathcal{A}_B = \partial_B \lambda$, where $\lambda$ is a gauge parameter.

We, therefore, get the radial fall-off conditions
\begin{align} \label{edAfall}
    A_u = \mathcal{O} (r^{-1}), &&  A_r = 0, && A_B = \mathcal{O}(r^0) .
\end{align}

To strengthen our argument, we check if the fall-off conditions \ref{edAfall} are consistent with the equations of motion.
To this end, we need to find the fall-off conditions of the coupled current $J_\mu$.

A reasonable condition to impose on the asymptotic behaviour of a general charge current $J_{\mu}$ is that the charge flux through $\mathcal{I}^+$ should be finite \cite{He2020}. Formally, this is expressed by
\begin{equation}
    | \int_{\mathcal{I}^+} J_{\mu} d\Sigma^{\mu} | < \infty .
\end{equation}
We obtain the large-$r$ fall-off condition
\begin{equation}
    J_u, J_r \in \mathcal{O}(r^{-2}) .
\end{equation}

However, we are particularly interested in currents generated by massless, under $U(1)$ symmetry charged fields, as they are sources for the electromagnetic field.

In the following, we will choose the complex massless scalar field $\phi$ to generate the current.
The massless scalar field has the following fall-off conditions at $r \rightarrow \infty$: \cite{Mitra2017, He2020}
\begin{equation} \label{scalfall}
    \phi, \phi^* \in \mathcal{O}(r^{-1}) .
\end{equation}

The Noether current following from the global $U(1)$ symmetry \ref{scacur} is given by $J_{\mu} = i(\nabla_{\mu} \phi^* \phi - \phi^* \nabla_{\mu} \phi)$ .

The fall-off conditions motivate an asymptotic expansion of the fields of the following form \cite{Mitra2017}
\begin{equation}
    \phi (u,r, \boldsymbol{\theta}) = \sum_{n=1}^{\infty} \frac{\phi^{(-n)} (u, \boldsymbol{\theta})}{r^n}. 
\end{equation}

To obtain the new fall-off conditions for the charge current we plug this into the equation for the Noether current, which leads to 
\begin{align} \label{edcurfall}
    J_u \in \mathcal{O}(r^{-2}), && J_r \in \mathcal{O}(r^{-4}), && J_A \in \mathcal{O}(r^{-2}) .
\end{align}

The $J_r \propto r^{-3}$ term vanishes as one can directly check \footnote{If one considers only the first order expansion of $\phi(u,r, \boldsymbol{\theta})$: $J_r =  i(\nabla_{r} \phi^* \phi - \phi^* \nabla_{r} \phi)  \approx i (- \frac{\phi^{*(-1)} (u, \boldsymbol{\theta})}{r^2} \frac{\phi^{(-1)} (u, \boldsymbol{\theta})}{r} + \frac{\phi^{*(-1)} (u, \boldsymbol{\theta})}{r} \frac{\phi^{(-1)} (u, \boldsymbol{\theta})}{r^2}) = 0$. The term $J_r \propto r^{-3}$ therefore vanishes and we therefore conclude that $J_r \in \mathcal{O}(r^{-4})$.} \cite{Schröder2022}.

Note, that we used a global $U(1)$ current to motivate the fall-off conditions. However, at the beginning of the chapter we stated, that the current must be invariant under local $U(1)$ gauge transformations, which doesn't apply to the current generated by the scalar fields.
To find the Noether current associated to a local $U(1)$ symmetry we use the gauge covariant derivative $\mathcal{D}_\mu \phi = \partial_\mu \phi - i A_\mu \phi$ and find $J_\mu = i ((\mathcal{D}_\mu \phi)* \phi - \phi^* \mathcal{D}_\mu \phi) = i \partial_\mu \phi^* \phi - i \phi^* \partial_\mu \phi - 2 \phi^* \phi A_\mu$ .
One can see that this current exhibits the same fall-off behaviour as in \ref{edcurfall} by inserting the fall-off conditions for $\phi$ \ref{scalfall} and $A_\mu$ \ref{edAfall}. 

To verify the consistency of the gauge field and current expansions, we consider the equations of motion for expansions of $n \geq 1$:
\begin{eqnarray}
    J_u^{(-2)} = \partial_u [\gamma^{AB} D_A A_B^{(0)} - A_u^{(-1)}] \\
    J_u^{(-n-2)}= - (-n-1) \partial_u A_u^{(-n-1)} - [n(n-1) + D^2] A_u^{(-n)} + \partial_u \gamma^{AB} D_A A_B^{(-n)} \\
    J_r^{(-n-3)} = -n(n+1) A_u^{(-n-1)} - n \gamma^{AB} D_A B_B^{(-n)} \\
    J_C^{(-2)} = -2 \partial_u A_C^{(-1)} + \partial_C A_u^{(-1)}- \gamma^{AB} D_A (D_B A_C^{(0)} -  D_C A_B^{(0)}) \\
    J_C^{(-n-2)} = (n+1) (-2 \partial_u A_C^{(-n-1)} + \partial_C A_u^{(-n-1)} - n A_C^{(-n)})  - \gamma^{AB} D_A D_{[B} A_{C]}^{(-n)}
\end{eqnarray}

Because the current falls off faster than the gauge fields, the fall-off conditions are consistent with the equations of motion.

\section{Extended Phase Space on Future Null Infinity}
The procedure to construct the covariant phase space has been laid out in chapter \ref{chap23}.
We raised the issue of the invertibility of the symplectic form in an infinite dimensional phase space, but didn't answer it, yet. We will demonstrate a solution to this problem in the following. This chapter is a summary of parts of \cite{Schröder2022, RojoSchröder2022}.

To construct the extended phase space on $\mathcal{I}^+$ , we first obtain the symplectic potential from the variation of the action \ref{edactionwo}:
\begin{equation}
    \Theta_\mu = -F_{\mu\nu} \delta A^\nu .
\end{equation}
We calculate the symplectic current with equation \ref{omega}
\begin{equation}
    \omega_\mu = \delta \Theta_\mu = - (\delta(\partial_\mu A_\nu) - \delta (\partial_\nu A_\mu)) \wedge \delta A^\nu .
\end{equation}

Using the fall-off conditions \ref{edAfall} and the directed volume element we determine the presymplectic form

\begin{equation} \label{edsymp}
    \begin{split}
        \tilde{\Omega}_{\mathcal{I}^+} &= \int_{\mathcal{I}^+} \omega_\mu d \Sigma^\mu = - \int_{\mathcal{I}^+} ((\delta(\partial_\mu A_\nu) - \delta (\partial_\nu A_\mu)) \wedge \delta A^\nu) d \Sigma^\mu \\
        &=\int_{\mathcal{I}^+} \gamma^{AB} \delta(\partial_u \mathcal{A}_A) \wedge \delta \mathcal{A}_B \ du  d\Gamma .
    \end{split}
\end{equation}

As before, we aim to find the Poisson brackets.
Now we arrived at the point where the problem of invertibility comes up. Naively constructing the Poisson brackets like in the example with the scalar field at a constant time in chapter \ref{chap23} would lead to inconsistencies, because the Poisson brackets are not well-defined \cite{Schröder2022, Mohd2015, Campiglia2014}. 
One deals with this problem by extending the phase space \cite{He2014}.

The first step is to rewrite the symplectic form and parametrize the extended phase space. 
The gauge field can be expressed as the sum of a $u$-dependent and $u$-independent part:
\begin{equation} \label{edsplit}
    \mathcal{A}_B (u, \boldsymbol{\theta}) = \hat{\mathcal{A}}_B (u, \boldsymbol{\theta}) + G_B(\boldsymbol{\theta}) ,
\end{equation}
where we defined
\begin{equation}
    G_B (\boldsymbol{\theta}) = \frac{1}{2} (\mathcal{A}^+_B (\boldsymbol{\theta}) + \mathcal{A}^-_B (\boldsymbol{\theta})) ,
\end{equation}
with the limits $\mathcal{A}^+_B(\boldsymbol{\theta}) = \lim_{u \rightarrow \infty} \mathcal{A}_B (u, \boldsymbol{\theta})$ and $\mathcal{A}^-_B(\boldsymbol{\theta}) = \lim_{u \rightarrow -\infty} \mathcal{A}_B (u, \boldsymbol{\theta})$ .

We furthermore define \footnote{$N_A$ is called the soft-photon field. It is a measure of how many soft photons go through $\mathcal{I}^+$ per angle. This can be derived by considering the Fourier transform of $\mathcal{F}_{uB} = \partial_u \mathcal{A}_B$ , which is the radiative part of the field strength: $FT \mathcal{F}_{uB} (\omega, \boldsymbol{\theta}) = \int_\mathbb{R} du \ e^{i\omega u} \partial_u \mathcal{A}_B (u, \boldsymbol{\theta})$. The soft photon field is then $N_B(\boldsymbol{\theta}) = FT\mathcal{F}_{uB} (0, \boldsymbol{\theta})$. \cite{Schröder2022, Miller2021}}
\begin{equation}
    N_B (\boldsymbol{\theta}) = \mathcal{A}^+_B (\boldsymbol{\theta}) - \mathcal{A}^-_B (\boldsymbol{\theta}) .
\end{equation}

We decomposed the gauge field into the pure gauge part $G_B (\boldsymbol{\theta})$ and a gauge-invariant time-dependent part $\hat{\mathcal{A}}_B (u, \boldsymbol{\theta})$.  $G_B (\boldsymbol{\theta})$ is pure gauge, because we imposed that there is no magnetic field at the boundaries $I^+_\pm$, i.e. $\mathcal{F}_{AB} = 0$ \cite{Strominger2018}. Note, that $\mathcal{A}^+_B (\boldsymbol{\theta}) =- \mathcal{A}^-_B (\boldsymbol{\theta})$. We will now treat these fields as independent.

Next, we insert the decomposition into the presymplectic form \ref{edsymp} \cite{Strominger2018}
\begin{equation}
    \tilde{\Omega}_{\mathcal{I}^+} = \int_{\mathcal{I}^+} \gamma^{AB} \delta (\partial_u \hat{\mathcal{A}}_A (u, \boldsymbol{\theta}) \wedge \delta \hat{\mathcal{A}}_B (u, \boldsymbol{\theta})) \ du d\Gamma  + \int_{S(u)} \gamma^{AB} \delta N_A (\boldsymbol{\theta}) \wedge \delta G_B (\boldsymbol{\theta}) d\Gamma .
\end{equation}

We parametrize the extended phase space by the field $\hat{\mathcal{A}}_B$ and the $u$-independent fields $G_B$ and $N_B$.

On this extended phase space we will now try to construct the Poisson brackets. First, we want to find $\{\hat{\mathcal{A}}_C (u, \boldsymbol{\theta}^{}), \hat{\mathcal{A}}_D (u^{\prime}, \boldsymbol{\theta}^{\prime})\}$. 

Like in the example for the scalar field, we try an ansatz of the form
\begin{equation}
    \begin{split}
        f = \int_{\mathcal{I}^+} \gamma^{AB} f_A (u, \boldsymbol{\theta}) \hat{\mathcal{A}}_B (u, \boldsymbol{\theta}) \ du  d\Gamma , \\
        g = \int_{\mathcal{I}^+} \gamma^{AB} g_A (u, \boldsymbol{\theta}) \hat{\mathcal{A}}_B (u, \boldsymbol{\theta}) \ du  d\Gamma .
    \end{split}
\end{equation}

The corresponding Hamiltonian vector fields on phase space are given by \\
$i_{X_f} \Omega_{\mathcal{I}^+} = -\delta f$ and $i_{X_g} \Omega_{\mathcal{I}^+} = -\delta g$  .

Recall, that the Poisson bracket between the arbitrary functions $f,g \in \mathfrak{F}(\Gamma)$ is defined as $\{f,g\} = - \Omega_{\mathcal{I}^+} (X_f, X_g)$. 

We start by determining the Hamiltonian vector field corresponding to $f$
\begin{equation}
    X_f = \int_{\mathcal{I}^+} x_{h;A} (u, \boldsymbol{\theta}) \frac{\delta}{\delta \hat{A}_A (u, \boldsymbol{\theta})} du  d\Gamma .
\end{equation}

To connect it with the ansatz for $f$, we calculate
\begin{equation} \label{e332}
    i_{X_f} \Omega_{\mathcal{I}^+} = 2 \int_{\mathcal{I}^+} \gamma^{AB} \partial_u x_{f;A} (u, \boldsymbol{\theta}) \ du  d\Gamma - \int_{S} \gamma^{AB} x_{f;A} (u, \boldsymbol{\theta}) \delta \hat{\mathcal{A}}_B (u, \boldsymbol{\theta}) |_{u=\pm \infty} d\Gamma 
\end{equation}
where we integrated by parts.

By definition, this has to be the same as
\begin{equation}
    - \delta f= - \int_{\mathcal{I}^+} \gamma^{AB} f_A (u, \boldsymbol{\theta}) \delta \hat{\mathcal{A}}_B (u, \boldsymbol{\theta}) \ du  d\Gamma .
\end{equation}

Now, we use the relation $\mathcal{A}^+_B (\boldsymbol{\theta}) =- \mathcal{A}^-_B (\boldsymbol{\theta})$, we obtained by extending the phase space.

Evaluating the second term of \ref{e332}  at $u=\pm \infty$, we obtain
\begin{equation}
    \begin{split}
        i_{X_f} \Omega_{\mathcal{I}^+} &= 2 \int_{\mathcal{I}^+} \gamma^{AB} \partial_u x_{f;A} (u, \boldsymbol{\theta}) \ du  d\Gamma \\
        &\mathrel{\phantom{=}} - \int_{S} \gamma^{AB} (x_{f;A}(\infty, \boldsymbol{\theta}) +  (x_{f;A}(- \infty, \boldsymbol{\theta}) ) \delta \hat{\mathcal{A}}_B^+ (\boldsymbol{\theta}) d\Gamma
    \end{split}
\end{equation}

and, therefore,
\begin{equation}
    \begin{split}
        i_{X_f} \Omega_{\mathcal{I}^+} &= 2 \int_{\mathcal{I}^+} \gamma^{AB} \partial_u x_{f;A} (u, \boldsymbol{\theta}) \ du  d\Gamma \\
        &\mathrel{\phantom{=}} - \int_{S} \gamma^{AB} (x_{f;A}(\infty, \boldsymbol{\theta}) +  (x_{f;A}(- \infty, \boldsymbol{\theta}) ) \delta \hat{\mathcal{A}}_B^+ (\boldsymbol{\theta}) d\Gamma \\
        &\stackrel{!}{=} - \delta f= - \int_{\mathcal{I}^+} \gamma^{AB} f_A (u, \boldsymbol{\theta}) \delta \hat{\mathcal{A}}_B (u, \boldsymbol{\theta}) \ du  d\Gamma .
    \end{split}
\end{equation}

As a result, we conclude that
\begin{equation}
    x_{f;A} = -\frac{1}{4} \gamma_{AC} sgn(u-u^\prime) \delta_{S} (\boldsymbol{\theta}, \boldsymbol{\theta}^\prime) ,
\end{equation}
introducing the sign-function with the property $\partial_u sng(u) = 2 \delta(u)$. This yields
\begin{equation}
    f_A = \gamma_{AC} \delta(u-u^\prime) \delta_{S} (\boldsymbol{\theta}, \boldsymbol{\theta}^\prime)
\end{equation}
and
\begin{equation}
    f= \int_{\mathcal{I}^+} \delta^B_D \delta(u-u^{\prime}) \delta_{S} (\boldsymbol{\theta}, \boldsymbol{\theta}^{\prime}) \hat{\mathcal{A}}_B (u, \boldsymbol{\theta}) \ du  d\Gamma .
\end{equation}

Analogously, we get
\begin{equation}
    g_A = \gamma_{AD} \delta(u-u^{\prime\prime}) \delta_{S} (\boldsymbol{\theta}, \boldsymbol{\theta}^{\prime\prime})
\end{equation}
and
\begin{equation}
    g= \int_{\mathcal{I}^+} \delta^B_D \delta(u-u^{\prime\prime}) \delta_{S} (\boldsymbol{\theta}, \boldsymbol{\theta}^{\prime\prime}) \hat{\mathcal{A}}_B (u, \boldsymbol{\theta}) \ du  d\Gamma .
\end{equation}
\cite{Mohd2015}

Finally, we determine the Poisson brackets
\begin{equation} \label{edpoisson1}
    \begin{split}
        \{\hat{\mathcal{A}}_C (u, \boldsymbol{\theta}^{}), \hat{\mathcal{A}}_D (u^{\prime}, \boldsymbol{\theta}^{\prime})\} &= i_{X_g} \Omega_{\mathcal{I}^+} (X_f) = - \delta g (X_f) \\
        &= - \frac{1}{4} \gamma_{CD} sgn(u - u^{\prime}) \delta_{S} (\boldsymbol{\theta}^{}, \boldsymbol{\theta}^{\prime} ) .
    \end{split}
\end{equation}

Because of the sign-function, the Poisson bracket is antisymmetric, as it should be.
Repeating this for the other Poisson brackets we obtain
\begin{equation} \label{edpoisson2}
    \{ N_C(\boldsymbol{\theta}^{}), G_D(\boldsymbol{\theta}^{\prime})\} = -\gamma_{CD} \delta_{S} (\boldsymbol{\theta}^{}, \boldsymbol{\theta}^{\prime}) .
\end{equation}
However, we recall that to make the symplectic form invertible, we assumed the fields $\hat A_A$, $G_A$ and $N_A$ to be independent. Recalling the definition of $N_A = \int \hat A_A du$, we see that this is not consistent. This was noticed in \cite{RojoSchröder2022} and is apparently because the covariant phase space method seems to be not suitable for null hypersurfaces.

We notice that the computation of the Poisson brackets was much easier in the scalar field example in chapter \ref{chap23}. This is because we considered a constant time slice instead of future null infinity. On an equal time slice, the fields and their time-derivatives are independent. This is not true anymore for $v$-hypersurfaces. Here, the field and the $u$-derivative are not independent anymore. Poisson brackets on a constant time slice at different spatial coordinates vanish, because they are not causally related. On the future null infinity hypersurface, spatial coordinates are causally related. While the angular components are causally unrelated at different $u, u^\prime$ and we get the expected delta function, the $u-u^\prime$ are taken into account by the sign-function. \cite{Schröder2022, RojoSchröder2022}

\section{Generating Charges of the Asymptotic Symmetries}

In this section, we will compute the generating charges of the asympotic symmetries.
The asymptotic symmetries are residual gauge transformations.

By imposing the retarded radial gauge $A_r = 0$, $A_u |_{\mathcal{I}^+} = 0$ we restriced the residual gauge parameter to be $\lambda = \lambda(\boldsymbol{\theta})$ at $\mathcal{I}^+$. The remaining symmetry of the theory at $\mathcal{I}^+$ is, therefore,
\begin{equation}
    \delta_\lambda \mathcal{A}_B (u, \boldsymbol{\theta}) = \partial_B \lambda (\boldsymbol{\theta}) .
\end{equation}
This is a transformation of the angular components of the leading order gauge field.

The generating charge of a transformation can be calculated with Noether's procedure as introduced in chapter \ref{chap21}.
The Noether current for a non-gauge fixed Maxwell theory, i.e. $\lambda = \lambda (u,r, \boldsymbol{\theta})$, is
\begin{equation} \label{edcur}
    j_\mu (\lambda) = J_\mu \lambda - g^{\nu\rho} F_{\mu\nu} \partial_\rho \lambda ,
\end{equation}
which is conserved $\nabla_\mu j^\mu (\lambda) = 0$.
The corresponding charge can be defined as
\begin{equation} \label{edchar}
    Q[\lambda] = \int_{\mathcal{I}^+} j_\mu (\lambda) d\Sigma^\mu .
\end{equation}

For vanishing gauge parameters at future null infinity, $\lambda |_{\mathcal{I}^+} = 0$ , the charge is zero. These are the so called "proper" gauge transformations. Asymptotic symmetries are those, which are non vanishing at infinity, also called "improper" gaug transformations. They are generated by non vanishing charges. To find them we already fixed a gauge and imposed boundary conditions at infinity. Now we only have to combine these information.

As stated above, fixing the gauge left us with $\lambda (u,r, \boldsymbol{\theta}) = \lambda (\boldsymbol{\theta})$.
Next, we plug this transformation into the Noether current \ref{edcur} and integrate it over $\mathcal{I}^+$ to get the conserved charge \ref{edchar}.

Staying in the extended phase space formalism and applying the fall-off conditions \ref{edAfall}, \ref{edcurfall}, we arrive at
\begin{equation}
    Q[\lambda] = \int_{\mathcal{I}^+} \mathcal{J}_u (u, \boldsymbol{\theta}) \lambda (\boldsymbol{\theta}) \ du  d\Gamma - \int_{S} \lambda (\boldsymbol{\theta}) D^A N_A(\boldsymbol{\theta}) \ d\Gamma .
\end{equation}

The asymptotic symmetries are actually only generated by the second term of the charge, called soft charge. The first term, called hard charge, arises because we coupled the fields to a current. \footnote{Per definition, the hard charge contains matter fields carrying energy, whereas the soft charge contains only fields of vanishing energy.}

What is left is to check if these charges really do generate the asymptotic symmetries.
To this end, we consider the fields  $\hat{\mathcal{A}}_B$, $G_B$ and $N_B$.

First, we apply the asymptotic symmetry $\lambda(\boldsymbol{\theta})$ on these fields.
The transformation is defined as
\begin{equation}
    \mathcal{A}_B (u, \boldsymbol{\theta}) \rightarrow \mathcal{A}_B (u, \boldsymbol{\theta}) + \partial_B \lambda(\boldsymbol{\theta}) .
\end{equation}
Because $\partial_B \lambda(\boldsymbol{\theta})$ is independent of $u$,  $G_B (\boldsymbol{\theta})$, which is also $u$-independent, must transform the same as $\mathcal{A}_B (u, \boldsymbol{\theta})$:
\begin{equation} \label{gaugeg}
    G_B (\boldsymbol{\theta}) \rightarrow G_B (\boldsymbol{\theta}) + \partial_B \lambda(\boldsymbol{\theta}) .
\end{equation}
From \ref{edsplit} we can see that
\begin{equation} \label{gaugeahat}
    \hat{\mathcal{A}}_B (u, \boldsymbol{\theta}) = \mathcal{A}_B (u, \boldsymbol{\theta}) - G_B(\boldsymbol{\theta}) \rightarrow \mathcal{A}_B (u, \boldsymbol{\theta}) + \partial_B \lambda(\boldsymbol{\theta}) - G_B(\boldsymbol{\theta})  - \partial_B \lambda(\boldsymbol{\theta}) = \hat{\mathcal{A}}_B (u, \boldsymbol{\theta})
\end{equation}
is invariant under these transformations.
Analogously, we see that 
\begin{equation} \label{gaugen}
    N_A (\boldsymbol{\theta}) \rightarrow N_A (\boldsymbol{\theta})
\end{equation}
is invariant, as well.

To check if our derived charge is really the generator of the asymptotic symmetry, we compute the Poisson brackets 
\ref{edpoisson1}, \ref{edpoisson2} \cite{Strominger2018, Schröder2022}
\begin{equation}
    \begin{split}
        \{\hat{\mathcal{A}}_A (u, \boldsymbol{\theta}), Q[\lambda]\} &= 0 \\
        \{N_B, Q[\lambda]\} &= 0 , \\
        \{G_B, Q[\lambda]\} &= \int_{S} \gamma^{CD} \partial_C \lambda ({\boldsymbol{\theta}^\prime}) \{ G_B(\boldsymbol{\theta}), N_D(\boldsymbol{\theta}^{\prime})\} \ d\Gamma \\
        &= \int_{S} \gamma^{CD} \partial_C \lambda ({\boldsymbol{\theta}^\prime}) \gamma_{BD} \delta_{S} (\boldsymbol{\theta}, \boldsymbol{\theta}^{\prime}) \ d\Gamma \\
        &= \partial_B \lambda (\boldsymbol{\theta}) .
    \end{split}
\end{equation}

This is what we expected because of \ref{gaugeahat}, \ref{gaugeg}, \ref{gaugen}.
One can also calculate the Poisson bracket of two charges and see that they obey the trivial $U(1)$-algebra 
$\{ Q[\lambda], Q[\lambda^\prime] \} = 0$.\cite{Schröder2022, Mitra2017}

\chapter{Asymptotic Symmetries of Kalb-Ramond Theory}
\label{chap4}
In the following, we will derive the generating charges of the asymptotic symmetries of the Kalb-Ramond theory.

The Kalb-Ramond theory is a generalization of Maxwell's theory of electrodynamics in which the gauge field is given by a two-form instead of a one-form \cite{Kalb1974}.

While in electrodynamics the gauge field $A_\mu$ couples to charged point particles, the Kalb Ramond field, denoted by $B_{\mu\nu}$, couples to two-dimensional strings in four spacetime dimensions. It allows us to describe "electrically charged" strings. \cite{Tong2009} 

An interesting property of the Kalb Ramond theory in four dimensions is that it is dual to a scalar field theory in four dimensions \cite{Weinberg2013}. This enables us to draw parallels between the two theories. We will therefore compare the fall-off conditions we get from the usual algorithm laid out in chapter \ref{chap3} with fall-off conditions we obtain from the duality relation. They will turn out to be consistent and might offer a new perspective to asymptotic symmetries in scalar field theories.

\section{Action and Variation}
The action for the free massless Kalb-Ramond theory is given by \cite{Kalb1974}
\begin{equation}
    S=\frac{1}{12} \int d^4x \sqrt{-g} H_{\mu\nu\rho} H^{\mu\nu\rho} .
\end{equation}

This is similar to the Maxwell action, but the two-form $F_{\mu\nu}$ and the one-form $A_\mu$ have been replaced by forms of one rank higher. $H_{\mu\nu\rho} =\partial_\mu B_{\nu\rho} + \partial_\rho B_{\mu\nu} + \partial_\nu B_{\rho\mu}$ is the field strength tensor for the Kalb-Ramond field $B_{\mu\nu}$, which is an antisymmetric rank-two tensor.
Like electrodynamics, the Kalb-Ramond theory is a gauge theory.

The field strength is gauge invariant under the the gauge transformation $\delta B_{\mu\nu} = \partial_\mu \epsilon_\nu - \partial_\nu \epsilon_\mu$. In contrast to electrodynamics, the gauge parameter $\epsilon_\mu$ is also subject to a gauge symmetry $\delta \epsilon_\mu = \partial_\mu f$, where $f$ is a scalar field.

Variation of the action leads to
\begin{equation}
    \delta S=-\int d^4x \sqrt{-g} (\nabla^\mu H_{\mu\nu\rho}) \delta B^{\nu\rho} + \int d^4x \sqrt{-g} \nabla^\mu (H_{\mu\nu\rho} \delta B^{\nu\rho}) .
\end{equation}

As usual, the bulk term gives us the equations of motion
\begin{equation}
    \nabla^\mu H_{\mu\nu\rho} = 0 .
\end{equation}

Until now, we considered the free Kalb-Ramond theory. Of course, we can couple the fields to a current, like in electrodynamics.
This current is a non-dynamical conserved and gauge invariant two-form $J^{\mu\nu}$. Coupled to this current, the action becomes
\begin{equation}
    S=\frac{1}{12} \int d^4x  \sqrt{-g}(H_{\mu\nu\rho} H^{\mu\nu\rho} + B_{\mu\nu} J^{\mu\nu})
\end{equation}
and by means of its variation
\begin{equation}
    \delta S=-\int d^4x \sqrt{-g} (\nabla^\mu H_{\mu\nu\rho} - J_{\nu\rho}) \delta B^{\nu\rho} + \int d^4x \sqrt{-g} \nabla^\mu (H_{\mu\nu\rho} \delta B^{\nu\rho})
\end{equation}

one obtains the equations of motion
\begin{equation}
    \nabla^\mu H_{\mu\nu\rho} = J_{\nu\rho}.
\end{equation}

\section{Radial Fall-off Conditions of the Field Strength}

The next step is to find reasonable large-$r$ fall-off conditions. Analogously to the electrodynamical case, we impose the condition that the energy, momentum and angular momentum fluxes through $\mathcal{I}^+$ has to be finite.

Again we construct the conserved corresponding current \ref{Killingcur} using the energy momentum tensor $T^{\mu\nu} = \frac{3}{2} H^{\mu\alpha\beta} H^\nu_{\alpha\beta} - \frac{1}{4} g^{\mu\nu} H^{\alpha\beta\gamma} H_{\alpha\beta\gamma}$ \cite{Belich2010} and the Killing vector field solution with the only non-zero component $X^r$ \cite{Schröder2022}.

Additionally, we also demand the charge flux at $\mathcal{I}^+$ to be finite $| \int_{S(u)} J_{\nu\rho} dS^{\nu \rho} | =  | \int_{S(u)} (\nabla^{\mu} H_{\mu \nu \rho}) dS^{\nu \rho} |   < \infty$ , where we integrated over the two-sphere $S(u)$ at a fixed retarded time $u$ and for large $r$.

From this condition and \ref{finitness}, we obtain the following fall-off conditions for the field strength (c.f. appendix \ref{appendixB1})
\begin{align}  \label{Hfall}
    H_{urA} \in \mathcal{O}(r^{-1}), &&  H_{uAB} \in \mathcal{O}(r), && H_{rAB} \in \mathcal{O}(r^{0}).
\end{align}

In 4 dimensions, the free Kalb-Ramond theory is dual to the scalar field theory \cite{Weinberg2013}. Thus, it could be interesting to apply this duality on the fields and find fall-off conditions we can compare to the ones we obtained above.
In 4 dimensions, the duality $H = dB = *d \phi$ holds on-shell, where $*$ denotes the Hodge star operator.

The standard fall-off condition for the massless scalar field from \ref{scalfall} are
\begin{equation}
    \phi (u,r,\boldsymbol{\theta}) = \frac{\phi^{(-1)} (u,\boldsymbol{\theta})}{r} + ...
\end{equation}

Using the duality relation and expanding the scalar fields asymptotically yields the following field strength fall-off behaviour (c.f. appendix \ref{appendixB2})
\begin{equation}
    \begin{split}
        H_{urA} = \frac{H^{(-1)}_{urA}}{r} + ... ,\\
        H_{uAB} = r H^{(1)}_{uAB} + ... ,\\
        H_{rAB} = H^{(0)}_{rAB} + ... . \\
    \end{split}
\end{equation}

Interestingly, this behaviour is consistent with the fall-off conditions \ref{Hfall} we derived above.

\section{Asymptotic Symmetries in "Radial" Gauge}

\subsection{Radial Fall-off Conditions of the Gauge Field}

In this section we want to find the fall-off conditions for the gauge fields $B_{\mu\nu}$.
First, we fix a gauge. We try to implement a "radial gauge", just like in electrodynamics.

We can always choose the $r$-dependence of $f(u,r,\boldsymbol{\theta})$ so that $\epsilon_r = 0$, giving $B_{\mu\nu} (u,r, \boldsymbol{\theta})$ , $\epsilon_\mu (u,r,\boldsymbol{\theta}) = (\epsilon_u, 0, \epsilon_A, \epsilon_B)$ and $f(u,\boldsymbol{\theta})$.
The remaining gauge  transformations are
\begin{equation} \nonumber
    \begin{split}
        B_{r\mu} &\rightarrow B_{r\mu} + \partial_r \epsilon_\mu \\
    B_{uA} &\rightarrow B_{uA} + \partial_u \epsilon_A - \partial_A \epsilon_u \\
    B_{AB} &\rightarrow B_{AB} + \partial_A \epsilon_B - \partial_B \epsilon_A \\
    \epsilon_u &\rightarrow \epsilon_u + \partial_u f \\
    \epsilon_A &\rightarrow \epsilon_A + \partial_A f
    \end{split}
\end{equation}

We can also choose the $r$-dependence of $\epsilon_\mu (u,r,\boldsymbol{\theta})$ so that $B_{r\mu} = 0$, giving the only remaining components of $B_{\mu\nu} (u,r,\boldsymbol{\theta})$: $B_{uA}, B_{AB}$,
 $\epsilon_\mu (u,\boldsymbol{\theta}) = (\epsilon_u, 0, \epsilon_A, \epsilon_B)$ and $f(u,\boldsymbol{\theta})$.
The remaining gauge  transformations are
\begin{equation} \nonumber
    \begin{split}
        B_{uA} &\rightarrow B_{uA} + \partial_u \epsilon_A - \partial_A \epsilon_u \\
        B_{AB} &\rightarrow B_{AB} + \partial_A \epsilon_B - \partial_B \epsilon_A \\
    \epsilon_u &\rightarrow \epsilon_u + \partial_u f \\
    \epsilon_A &\rightarrow \epsilon_A + \partial_A f
    \end{split}
\end{equation}

We may further choose the $u$-dependence of $f(u,\boldsymbol{\theta})$ so that $\epsilon_u  = 0$, giving
$B_{uA} (u,r, \boldsymbol{\theta})$,  $B_{uB} (u,r, \boldsymbol{\theta})$, $B_{AB} (u,r, \boldsymbol{\theta})$, $\epsilon_\mu (u,\boldsymbol{\theta}) = (0, 0, \epsilon_A, \epsilon_B)$ and $f(\boldsymbol{\theta})$.
The final residual gauge transformations are therefore
\begin{equation} \nonumber
    \begin{split}
        B_{uA} &\rightarrow B_{uA} + \partial_u \epsilon_A \\
        B_{AB} &\rightarrow B_{AB} + \partial_A \epsilon_B - \partial_B \epsilon_A \\
        \epsilon_A &\rightarrow \epsilon_A + \partial_A f
    \end{split}
\end{equation}

To check this gauge fixing, we consider the degrees of freedom. \footnote{Here, we mean the off-shell degrees of freedom (compared to on-shell/propagating degrees of freedom). They can be calculated by substracting the number of gauge transformations from the number of components. Generally, for p-forms in d dimensions, therer are $\binom{d-1}{p}$ off-shell degrees of freedom. The number of components is $\binom{d}{p}$ from which we substract the gauge transformations $\binom{d-1}{p-1}$. \cite{Afshar2018}}
The parameter $\epsilon$ has $3 = 4-1$ independent degrees of freedom, where one was removed by $f$. The gauge field $B$ has $3= 6-3=6-(4-1)$ independent degrees of freedom, where three were removed by $\epsilon$. The remaining components of $B$ exactly match the $3=4-1$ independent components of $H$, where one was removed by the Bianchi identity.

In this gauge, the equations of motion become in retarded Bondi coordinates
\begin{equation} \label{kreomrad}
\begin{split}
    J_{ur} = - \partial_r \gamma^{AB} D_B B_{Au} ,\\
    J_{uA} = (\partial_r^2 - \partial_u \partial_r) B_{uA} + \frac{1}{r^2} \partial_u D^B B_{AB} + \frac{D^2 - 1}{r^2} B_{uA} + \frac{1}{r^2} \gamma^{BC} D_A D_C B_{Bu} ,\\
    J_{rA}= -\partial_r^2 B_{uA} - \frac{1}{r^2} \partial_r \gamma^{BC} D_C B_{AB} ,\\
    J_{AB} = (2(\frac{1}{r} (\partial_u + \partial_r) - \partial_r \partial_u) - \partial^2_r+ \frac{D^2}{r^2}) B_{AB} + (\partial_r - \frac{2}{r}) D_{[A}B_{B]u} \\
    + \frac{1}{r^2} D_{[A} D^{C} B_{B]C} .
\end{split}
\end{equation}

From the field strenght fall-off conditions and the equations of motion, we obtain the the fall-off behaviour
\begin{align} \label{Bradfallplain}
     B_{AB} \in \mathcal{O}(r) , &&  B_{uA} \in \mathcal{O} (log(r)).
\end{align}

Furthermore, we can express $B_{uA}$ in terms of $B_{AB}$ because $B^{(1)}_{AB}$ is the only  propagating degree of freedom of the Kalb-Ramond field. \footnote{The propagating degrees of freedom (or on-shell degrees of freedom) are those which do not depend on the fields source and can therefore propagate freely. The number of propagating degrees of freedom for a p-form in d dimensions is $\binom{d-2}{p}$. \cite{Afshar2018}} The relation is given by the equations of motion and leads to
\begin{equation} \label{Bradfall}
    \begin{split}
        B_{AB} = r B_{AB}^{(1)} + ... \ , \\ 
        B_{uA} = log(r) D^B B_{AB}^{(1)}  + ... \ .
    \end{split}
\end{equation}

Again, we consider the dual scalar theory. Rewriting the components with the duality relation and inserting the fall-off conditions of the scalar field yields (c.f. appendix \ref{appendixB2})
\begin{align} \label{Bradfalldual}
     B_{AB} \in \mathcal{O}(r) , &&  B_{uA} \in \mathcal{O} (log(r)),
\end{align}

which is consistent with \ref{Bradfallplain}.

To verify the derived fall-off conditions, we check the consistency of the fall-off conditions with the equations of motion and the fall-off behaviour of the current. This was also done in the electrodynamical case. The current that couples to the two-form Kalb-Ramond field is a two-form as well.

To find the large-$r$ behaviour of the current, we demand a finite charge flux through future null infinity. Because $J^{\mu\nu}$ is a two-form we need to integrate over the two-sphere $S(u)$ at a fixed retarded time $u$ and for large $r$.
The conserved charge is defined as \cite{Afshar2018}
\begin{equation}
    Q = \int_{S(u)} * J .
\end{equation}
The oriented surface element for $S(u)$ is $dS_{AB}$ \cite{Afshar2018}.
The condition of the charge flux to be finite is therefore
\begin{equation}
    | \int_{S(u)} (* J)^{\mu\nu} d S_{\mu\nu}| = | \int_{S(u)}  (* J)^{AB} d S_{AB}| =| \int_{S(u)} J^{ur} \ r^2 d\Gamma| <\infty ,
\end{equation}
where the factor $r^2$ came from the definition of the Hodge star operator. The Hodge star operator also relates the $A,B$ components of $(* J)$ with the $u,r$ components of $J$. 

This condition imposes the fall-off behaviour
$J^{ur} \in \mathcal{O}(r^{-2})$ . This is the only restriction for components of $J^{\mu\nu}$ we obtain from the requirement of the finiteness of the charge flux at future null infinity.

Nevertheless, the condition we found is consistent with the fall-off behaviour of the fields as one can see by inserting the asymptotic expansions of $B$ and $J$ into the equations of motion \ref{kreomrad}. The current falls off faster than the gauge fields at infinity.

\subsection{Generating Charges}

The residual gauge transformations are $\epsilon_A (u,\boldsymbol{\theta})$. We want to compute the generating charges of these symmetries.
The charge is defined as \cite{Avery2016}
\begin{equation}
    Q^+ = \int_{\mathcal{I}^+} * j =  \int_{S(u)} *\kappa 
\end{equation}
where $\kappa^{\mu\nu}$ is the Noether two-form defined by the Noether current $j_\mu (\epsilon) = \nabla^\nu \kappa_{\mu\nu}$.
In components, we can write
\begin{equation}
    Q^+ = \int_{S(u)} \kappa^{ur} r^2 d\Gamma ,
\end{equation}
where we used that the Hodge star operator $*$ relates the $A,B$ components of $(* \kappa)$ with the $u,r$ components of $\kappa$ and integrate over the oriented surface element $dS_{AB}$. The Hodge star operator also causes the factor $r^2$.

The Noether two-form for the Kalb-Ramond action is $\kappa^{\mu\nu} = H^{\nu\mu\rho} \epsilon_\rho$ \cite{Avery2016}. Inserting the residual symmetry, this yields
\begin{equation}
    \kappa^{ur} = H^{ruA} \epsilon_A = g^{AB} H_{ruB} \epsilon_A = - \frac{1}{r^2} \gamma^{AB} \partial_r B_{Bu} \epsilon_A .
\end{equation}

Now, we can compute the generating charge of the asymptotic symmetries in radial gauge:
\begin{equation}
    \begin{split}
        Q^+ &= \int_{S(u)} \kappa^{ur} r^2 d\Gamma = - \int_{S(u)} \gamma^{AB} \partial_r B_{uB} \epsilon_A = - \frac{1}{r} \int_{S(u)} \gamma^{AB} \gamma^{CD}  \epsilon_A D_D B^{(1)}_{BC} d\Gamma \\
        &= \frac{1}{r} \int_{S(u)} \gamma^{AB} \gamma^{CD}  B^{(1)}_{BC} D_D \epsilon_A d\Gamma = \frac{1}{r} \int_{S(u)} \gamma^{AB} \gamma^{CD} B^{(1)}_{BC} (\partial_D \epsilon_A - \partial_A \epsilon_D)  d\Gamma
    \end{split}
\end{equation}

In the third equality we used \ref{Bradfall} and in the fourth we integrated by parts. Note, that this is the soft charge that generates the asymptotic symmetries. The Noether two-form was obtained without coupling the fields to a current, so we do not have a hard charge.

One can see that the computed generating charge vanishes at large $r$.
This is because in radial gauge, the gauge parameter $\epsilon_A$ does not scale like $r$ and therefore can not compensate for $\partial_r B_{uA} \sim \frac{1}{r}$.
We will comment on this result in section 4.5.

\section{Asymptotic Symmetries in Lorenz Gauge}

\subsection{Radial Fall-off Conditions of the Gauge Field}

In the following we will perform the same analysis as above, but this time in Lorenz gauge.
As we will see, the charge obtained by this procedure does not vanish at large $r$.

The condition imposed by the Lorenz gauge is
\begin{equation}
    \nabla^{\mu} B_{\mu\nu} = 0 .
\end{equation}
The residual gauge transformations are those $\epsilon_\mu$, that satisfy $\nabla^\mu (\partial_\mu \epsilon_\nu - \partial_\nu \epsilon_\mu) =0$. If we define the two-form $\eta = d\epsilon$, $\eta$ satisfies $d\eta = 0$ and $\nabla^\mu \eta_{\mu\nu}$. The large-$r$ behaviours of $\eta_{\mu\nu}$ must be the same or faster than those for $B$.

From the fall-off conditions derived above \ref{Hfall} and the equations of motion
\begin{align} \nonumber
    H_{urA} \in \mathcal{O}(r^{-1}),  && H_{uAB} \in \mathcal{O}(r), && H_{rAB} \in \mathcal{O}(r^{0}),
\end{align}
we conclude that the gauge fields must satisfy:
\begin{align} \label{Blorfall}
    B_{ur} \in \mathcal{O} (r^{-1}), && B_{uA} \in \mathcal{O} (r^{0}), && B_{rA} \in \mathcal{O} (r^{-1}), && B_{AB} \in \mathcal{O} (r) .
\end{align}

Again, these fall-off conditions are consistent with the ones obtained by duality considerations (c.f. appendix \ref{appendixB2}).
 
From the equations of motion in Lorenz gauge, we obtain the relation
\begin{equation} \label{rel}
    \partial_u B_{rA}^{(-1)} + D_A B_{ur}^{(-1)} = D^B B_{AB}^{(1)} ,
\end{equation}
which is going to be useful for expressig the generating charges using the only propagating degree of freedom.

\subsection{Generating Charges}
We can now calculate the generated charge the same way as in the "radial" gauge.
The Noether form in Lorenz gauge is expressed as 
\begin{equation}
    \kappa^{ur}_L = H^{ruA} \epsilon_A = g^{AB} H_{urB} \epsilon_A = -\frac{1}{r^2} \gamma^{AB} H_{urB} \epsilon_A .
\end{equation}

The charge is therefore
\begin{equation}
    \begin{split}
        Q^+ &= \int_{S(u)} \kappa^{ur}_L r^2 d\Gamma = -\int_{S(u)} \gamma^{AB} \frac{H_{urB}^{(-1)}}{r} \epsilon_A d\Gamma \\
        &= -\int_{S(u)} \gamma^{AB} \frac{1}{r}(\partial_B B_{ru}^{(-1)} + \partial_u B_{rB}^{(-1)}) \epsilon_A d\Gamma \\
        &= -\int_{S(u)} \gamma^{AB} \gamma^{CD} \frac{1}{r} D_D B_{BC}^{(1)} \epsilon_A d\Gamma \\
        &= \int_{S(u)} \gamma^{AB} \gamma^{CD} \frac{1}{r} B_{BC}^{(1)} D_D \epsilon_A d\Gamma = \int_{S(u)} \gamma^{AB} \gamma^{CD} B_{BC}^{(1)} \eta_{AB}^{(1)} d\Gamma \\
        &=\int_{S(u)} \gamma^{AB} \gamma^{CD} B_{BC}^{(1)} (\partial_D \epsilon_A - \partial_A \epsilon_D) d\Gamma ,
    \end{split}
\end{equation}
where we used the relation \ref{rel} and integrated by parts.

This charge does not vanish at infinity and looks very similar to the one obtained in the "radial" gauge. It appears that the Lorenz gauge might be a better choice in this case, as the radial gauge might be too strict. We will comment on this in section 4.5.

\section{Comments on the Charges, Covariant Phase Space and Duality}

We derived the generating charges in radial and Lorenz gauge and saw that the charge vanishes at infinity in radial gauge. We cross checked this result by verifying the fall-off conditions of the gauge fields with the fall-off conditions of the coupled current in the equations of motion.
Curiously, the charge in Lorenz gauge is the same charge as in radial gauge, but without the $\frac{1}{r}$ prefactor.
This is actually not totally unexpected, as different fall-off conditions and gauges can certainly lead to different conserved charges.
Fall-off conditions could, for example, be too strict and prohibit some residual asymptotic symmetries. At the same time, they can also be too loose so that the charge still generates some trivial gauge transformations, that should vanish at infinity.
By fixing different gauges, we indirectly impose different fall-off conditions. They might look the same, but have a different meaning, as the components after a gauge fixing also have a different meaning.
Therefore, the correspondence of two asymptotic charges in two different gauges is not warranted.
A cause for the charge in radial gauge to vanish at infinity might be that the conditions in radial gauge are too strict.
Also, the meaning of a radial gauge for extended objects is typically difficult to depict and not fully understood.

Just like in electrodynamics, one could proceed by constructing the covariant phase space at future null infinity but this would go beyond the scope of this thesis. Instead we will sketch the procedure. Starting from reading off the symplectic potential $\Theta_\mu = H_{\mu\nu\rho} \delta B^{\nu\rho}$, one would calculate the symplectic current $\omega_\mu = \delta \Theta_\mu = (\delta(\partial_\mu B_{\nu\rho} )+ \delta(\partial_\rho B_{\mu\nu}) + \delta(\partial_\nu B_{\rho\mu})) \wedge \delta B^{\nu\rho}$ .
This leads to the presymplectic form $\tilde\Omega_{\mathcal{I}^+} =  \int_{\mathcal{I}^+} \omega_\mu d \Sigma^\mu$, where one inserts the fall-off conditions.
Like electrodynamics, the Kalb-Ramond theory is also a gauge theory with an infinite dimensional phase space.
Therefore, one needs to extend the phase space. 
The first step would be to decompose the gauge field into a pure gauge part, the soft field, and the remaining gauge-invariant time-independent part. The decomposed fields are used to parametrize the phase space with. This was done in electrodynamics for $G_B(\boldsymbol{\theta})$ and $N_B(\boldsymbol{\theta})$ \ref{edsplit}. Then one can rewrite the presymplectic form. In the following the Poisson brackets could be constructed following the procedure in chapter \ref{chap3}.

At last, we will comment on the very interesting topic of duality.
Recently, soft charges for scalar field theory have been unveiled \cite{Campiglia2018}.
Therein, the authors wondered about which asymptotic symmetries can be applied to a scalar field.
Starting from the established soft theorem for scalar charges, they used connections in the infrared triangle, particularly the Ward identity, to calculate conserved charges that follow a similar structure like in the gauge field case. But they struggled to find the actual asymptotic symmetry associated with these charges.
In this chapter we observed that there is a consistency in the asymptotic behaviours of Kalb-Ramond theory and scalar field theory. Even after asymptotic expansion, the duality was compatible.
From the Kalb-Ramond theory perspective the asymptotic symmetries are much clearer.
Using the duality, in the future, one can attempt to relate the charges found in the scalar field case to the asymptotic symmetries of the Kalb-Ramond theory in order to explicitly find asymptotic symmetries for the scalar field and understand what such symmetries would actually mean.

\chapter{Summary and Conclusion}
\label{chap5}
Let us elaborate on some issues, as well as other points we consider important. We will also give suggestions for future research.

We started the thesis by presenting the concept of symmetries, Minkowski spacetime at future null infinity and the covariant phase space formalism.

Then, we reviewed the asymptotic symmetries of a rank-1 gauge field at future null infinity in 4 dimensional Minkowski spacetime. We carefully chose reasonable boundary conditions by demanding finiteness of the energy, momentum and angular momentum flux through future null infinity. After fixing the radial gauge, we expanded the gauge fields asymptotically and derived their large-$r$ behaviour. We further motivated these fall-off conditions by coupling the fields to a current, demanding its charge flux through future null infinity to be finite and checking the consistency of the fall-off conditions using the equations of motion.
Furthermore, we restricted the theory by excluding magnetic charges, which let the Coulombic order of radial magnetic fields vanish.
The choice of boundary conditions is very important, because too strict conditions would exclude physically essential solutions and too loose conditions would lead to divergences in physical quantities like the charge flux.

Next, we reviewed the construction of the phase space at future null infinity, for electromagnetism. We extended the phase space to ensure the invertibility of the symplectic form. The invertibility is needed to ultimately find consistent Poisson brackets. The extension was accomplished by introducing new fields that follow from decomposing the gauge field into a pure gauge part and a gauge-invariant time-dependent part. These fields were assumed to be independent and used to parametrize the extended phase space. 

Using Noether's formalism we computed the conserved charges of the asymptotic symmetries.

In the original part of this thesis, we determined the asymptotic symmetries of the Kalb-Ramond field at future null infinity. Kalb-Ramond theory is a generalization of Maxwell's electrodynamics and describes two-dimensional charged strings instead of point charges. The gauge field is now an antisymmetric two-form.
Once again, we obtained the field strength fall-off conditions by demanding energy and charge fluxes to be finite at future null infinity. Next, we tried to fix a "radial gauge" and expanded the gauge field in this gauge. The charge obtained by the remaining symmetries turned out to vanish at infinity.
In order to determine if this was caused by too strong boundary conditions, we coupled the field to a conserved two-current and compared its fall-off behaviours with those of the gauge field. We did not find any inconsistency.

Additionally, we repeated the process using the Lorenz gauge. This lead to a very similar but non-vanishing charge at future null infinity, which showcased the influence different gauges can have on the meaning of the boundary conditions and subsequently the remaining asymptotic symmetries.

An interesting point followed from the duality of Kalb-Ramond theory and scalar field theory in 4 dimensions.
We noticed the compatibility of the asymptotic behaviours of Kalb-Ramond theory and scalar field theory.
This could be a possible solution to the problem of asymptotic symmetries in scalar field theory. 
While asymptotic symmetries for scalar fields remain elusive, there are associated soft charges, derived from soft theorems \cite{Campiglia2018}. 
This is a research line which we highly encourage to further investigate.
It might be possible that the asymptotic symmetries for the Kalb-Ramond field explored herein can help to find and understand the corresponding asymptotic symmetries for the scalar field.

\appendix
\chapter{Metric and Differential Forms}
\section{Metric} \label{appendixmetric}
Herein we list the explicit form of Christoffel symbols in retarded Bondi coordinates on Minkowski spacetime.
The metric, its inverse and the determinant are

\begin{align}
     g_{\mu\nu} = \begin{pmatrix} 
                    1 & 1 & 0\\
                    1 & 0 & 0 \\
                    0 & 0 & -\frac{r^2}{\gamma^{AB}}
                    \end{pmatrix} ,
        && g^{\mu\nu} = \begin{pmatrix} 
                    0 & 1 & 0\\
                    1 & -1 & 0 \\
                    0 & 0 & -\frac{\gamma^{AB}}{r^2}
                    \end{pmatrix} ,
        && g = det(g_{\mu\nu}) = r^4 det(\gamma_{AB}).
\end{align}

The non-zero Christoffel symbols are given by
\begin{align}
    \Gamma^{u}_{AB} = r \gamma_{AB} = -  \Gamma^{r}_{AB} , && \Gamma^{A}_{rB} = \frac{1}{r} \delta^{A}_{B},  \\
     \Gamma^{\theta}_{\phi \phi} = sin(\theta) cos(\theta), && \Gamma^{\phi}_{\theta \phi} = \frac{1}{tan(\theta)}.
\end{align}
an the spherical parts satisfy $\Gamma^{A}_{BC} (g_{\mu\nu}) = \Gamma^{A}_{BC} (\gamma_{\mu\nu})$.

The covariant derivatives can be calculated via
\begin{equation}
    \nabla_\mu F_{\alpha\beta} = \partial_\mu F_{\alpha\beta} - \Gamma^\gamma_{\mu\alpha} F_{\gamma\beta} - \Gamma^\gamma_{\mu\beta} F_{\alpha\gamma},
\end{equation}
\begin{equation}
    \nabla_\mu H_{\alpha \beta \gamma} = \partial_\mu H_{\alpha \beta \gamma} - \Gamma^\rho_{\mu \alpha} H_{\rho \beta \gamma} - \Gamma^e{}_{\mu \beta} H_{\alpha \rho \gamma} - \Gamma^\rho_{\mu \gamma} H_{\alpha \beta \rho}
\end{equation}

\section{Differential forms} \label{appendixforms}
In this thesis, we frequently used differential forms and operators thereon like inner product or the Hodge star operator. Therefore, we will provide a short summary of the relevant concepts in this appendix based on \cite{Delft}.

Each vector space $V$ has a partner space called its dual space $V^*$. The dual space is defined as the set of all linear maps $w$ of $V$ into the real numbers
\begin{equation}
    w: V \rightarrow \mathbb{R}, v \mapsto w(v) \equiv wv .
\end{equation}
A linear map $w$ satisfys the properties $w(u+v) = wu + wv$ and $w(av)=awv$.

The vector space $V$ is spanned by the basis vectors $\{e_1, ..., e_n\}$ and the dual space by the dual basis vectors $\{e^1, ..., e^n\}$, which satisfy $e^i(e_j) = \delta^i_j$.
After choosing a basis, the vector's components have superscript indices and the components of linear maps have subscript indices:
\begin{align}
    v = v^i e_i ,  && w=w_i e^i .
\end{align}

Objects with superscript indices are called contravariant and those with subscript indices are called covariant.
To relate vectors and dual vectors one introduces a metrig $g$. A metric is a bilinear form, which means it maps two input vectors linearly to the real number:
\begin{align}
    g : V \times V \rightarrow \mathbb{R}, &&  v,u \mapsto g(v,u) = v \cdot u ,
\end{align}
or representing the vectors in their basis $v=v^i e_i, u = u^j e_j$ :
\begin{equation}
     g(v,u) = v^i u^i g(e_i, e_j) = v^i u^i e_i \cdot e_j = g_{ij} v^i u^j .
\end{equation}

The relation between vectors and their dual is given by
$v= v_i e^i = g(v, \cdot) = g_{ij} v^i e^j \Rightarrow v_i = g_{ij} v^i$ 
Therefore, index lowering or raising through the metric is equivalent to passing from a vector space to its dual vector space or back, in a component representation.

A tensor product $\otimes$ is a product of two vectors with the properties:
\begin{equation}
    \begin{split}
        a \otimes v = a v, \\
        (v+v^\prime) \otimes (u + u^\prime) = v \otimes u + v^\prime \otimes u + v \otimes u^\prime + v^\prime \otimes u^\prime , \\
        (av) \otimes u = v \otimes (au) = a (v \otimes u) ,
    \end{split}
\end{equation}
where $v, v^\prime, u, u^\prime \in V$ and $a \in \mathbb{R}$.

A $(p,q)$-tensor $T$ of contravariant rank $p$ and covariant rank $q$, is a multilinear map that maps $q$ vectors $v_i \in V$ and $p$ dual vectors $w^i \in V^*$ to the real numbers $\mathbb{R}$. $T$ is linear in every argument:
\begin{equation}
    \begin{split}
        T: \underbrace{V^* \times ... \times V^*}_{\textrm{p times}}  \times \underbrace{V \times ... \times V}_{\textrm{q times}} \rightarrow \mathbb{R} , \\
        T \mapsto T(w^1,...,w^p, v_1, ..., v_q) \in \mathbb{R} .
    \end{split}
\end{equation}
A tensor can be expressed with basis vectors
\begin{equation}
    T=T^{i_1 ... i_p}_{j_1 ... j_q} e_{i_1} \otimes ... \otimes  e_{i_p} \otimes e^{j_1} \otimes ... \otimes  e^{j_q} .
\end{equation}
If the components of the tensor are invariant under permutation of two indices it is called symmetric with respect to these indices, e.g. $T^{ij} = T^{ji}$. It is called antisymmetric if $T^{ij} = - T^{ij}$. A tensor that is antisymmetric with respect to all its indices is alternating.

Alternating forms $\phi$ with the covariant rank $p$ are called alternating multilinear forms or $p$-forms. They map $p$ vectors to the real numbers $\mathbb{R}$ and satisfy $\phi(..., v, ..., u,...) = -\phi(...,u,...,v,...)$.

Two multilinear forms $\phi$ of rank $p$ and $\psi$ of rank $q$ can be multiplied by the wedge product $\wedge$. Their product $(\phi \wedge \psi)$ is then a $(p+q)$-form.
The wedge product is bilinear, respects the associative and the graded commutative property ($\phi \wedge \psi = (-1)^{pq} \psi \wedge \phi$).
Therefore, $\phi \wedge \phi = 0$.
The wedge product of the dual basis vectors $\{e^1, ..., e^p\}$ can be expressed as
\begin{equation}
    e^i \wedge ... \wedge e^p = \epsilon_{i_1 ... i_p} e^1 \otimes ... \otimes e^{i_p} ,
\end{equation}
where $\epsilon_{i_1 ... i_p}$ is the antisymmetric Levi-Civita symbol.

A $p$-form in basis representation is therefore
\begin{equation}
    \phi = \frac{1}{p!} \phi_{i_1 ... i_p} e^i \wedge ... \wedge e^{i_p} .
\end{equation}
Because of the antisymmetry, a $p$-form has $\binom{n}{p}$ independent compoinents $\phi_{i_1 ... i_p}$, where $n$ is the dimension of the dual vector space where the dual basis vectors $e^i$ live.
If $p=n$, the form is called a topform and posesses ony one free component $\phi_{1...n}$. One can see that all forms of $p>n$ vanish.

The binomial coefficient satisfies the relation $\binom{n}{p} = \binom{n}{n-p}$. Therefore, a $p$-form has the same number of coefficients as a $(n-p)$-form. It should be possible to define a map between those forms. Indeed, this map is called the Hodge-star operator $*$.
Applying the Hodge-star operator on the dual basis vectors yields
\begin{equation}
    *( e^i \wedge ... \wedge e^{i_p}) = \frac{det(g)}{(n-p)!} \epsilon_{j_1 ... j_p j_{p+1} ... j_n} g^{i_1 j_1} g^{i_p j_p} e^{j_{p+1}} \wedge ... \wedge e^{j_n} .
\end{equation}
Starting from the $p$-form $\phi$ the $(n-p)$-form $*\phi$ is calculated by
\begin{equation}
    *\phi = *(\frac{1}{p!} \phi_{i_1 ... i_p} e^i \wedge ... \wedge e^{i_p}) = \frac{det(g)}{p!(n-p)!} \phi^{j_1 ... j_p}\epsilon_{j_1 ... j_p j_{p+1} ... j_n} e^{j_{p+1}} \wedge ... \wedge e^{j_n} .
\end{equation}
Note, that $** \phi = sign(g) (-1)^{p(n-p)} \phi$.

After choosing a coordinate system $(x^1,...,x^n)$ one can write a $p$-form as a differential form as a wedge product of the differentials $dx^i$:
$\phi = \frac{1}{p!} \phi_{i_1 ... i_p}(x) dx^i \wedge ... \wedge dx^{i_p}$, where $\phi_{1...p} = \phi(e_1,...,e_p)$.
The differential forms live on a 
Zero forms are simply functions $f(x)$ on a manifold.

Just like for a differential $df$ of a function or 0-form, we want to define a differential operator $d$ on a $p$-form $\phi$. The so-called exterior derivative $d$ maps a $p$-form to a $(p+1)$-form:
\begin{equation}
    d\phi = d (\frac{1}{p!} \phi_{i_1 ... i_p}(x) dx^i \wedge ... \wedge dx^{i_p}) = \frac{1}{p!} \frac{\partial \phi_{i_1 ... i_p}}{\partial x^j} dx^j \wedge dx^i \wedge ... \wedge dx^{i_p} .
\end{equation}

This object is called a differential form and is defined on a differentiable manifold \footnote{A differentiable manifold is locally similar enough to a vector space to apply calculus.}. The differential forms live in the cotangent space, which is the dual space to the tangent space. The tangent space is defined at a point $x$ and consists of all tangents starting from that point, for example on a curve.
Note, that top-forms have a vanishing exterior derivative and  $d \cdot d = 0$ .

After defining a map that connects $p$-forms $\phi$ and $(p+1)$-forms, we want to define a map that connects $p$-forms and $(p-1)$-forms. This map is called the interior product $i_v$:
\begin{equation}
    (i_v \phi) (v_1, ..., v_{p-1}) = \phi (v, v_1, ..., v_{p-1}) .
\end{equation}

It describes the process of "plugging a vector $v$ into a $p$-form".
The interior is linear
\begin{equation}
    i_{v+u} = i_v + i_u ,
\end{equation}
antisymmetric
\begin{equation}
    i_v \cdot i_u = - i_u \cdot i_v \Rightarrow i_v^2 = 0 ,
\end{equation}
and satisfys the graded product rule
\begin{equation}
    i_v (\phi \wedge \psi) = (i_v \phi) \wedge \psi + (-1)^p \phi \wedge (i_v \psi) .
\end{equation}

\chapter{Explicit Calculations}

\section{Fall-offs of the Kalb-Ramond Field Strengths} \label{appendixB1}
We start with the condition imposed on the energy, momentum and angular momentum flux. Recall \ref{finitness}: 
\begin{equation}
    |\int_{\mathcal{I}^+} r^2 (T_{ur} - \frac{1}{2} T_{rr}) |^2 < \infty .
\end{equation}
First, we will compute the components of the energy-momentum tensor \\
$T^{\mu\nu} = \frac{3}{2} H^{\mu\alpha\beta} H^\nu_{\alpha\beta} - \frac{1}{4} g^{\mu\nu} H^{\alpha\beta\gamma} H_{\alpha\beta\gamma}$ : \cite{Belich2010}
\begin{equation}
    \begin{split}
        T_{rr} &= g_{\mu r} g_{\nu r} T^{\mu \nu} = g_{ur} g_{ur} T^{uu} = T^{uu}= \frac{3}{2} H^{u\alpha \beta} g^{u a} H_{a \alpha \beta} \\
        &= \frac{3}{2} (H^{urA} g^{u u} H_{urA} + H^{uAB} g^{u u} H_{uAB}) \\
        &= \frac{3}{2} (g^{AB} H_{ruB} H_{ruA} + g^{AB} g^{BC} H_{rBC} H_{uAB} ) \\
        &= \frac{3}{2} (H^{urA} g^{u u} H_{urA} + H^{uAB} g^{u u} H_{uAB}) \\
        &= \frac{3}{2} (\frac{1}{r^2} \gamma^{AB} H_{ruB} H_{ruA} + \frac{1}{r^4} \gamma^{AB} \gamma^{BC} H_{rBC} H_{uAB} )
    \end{split}
\end{equation}

\begin{equation}
    \begin{split}
        T_{ur} &= g_{u\mu} g_{r \nu} T^{\mu\nu} = g_{uu} g_{ru} T^{uu} + g_{ur} g_{ru} T^{ur} \\
        &= \frac{3}{2} (\frac{1}{r^2} \gamma^{AB} H_{ruB} H_{ruA} + \frac{1}{r^4} \gamma^{AB} \gamma^{BC} H_{rBC} H_{uAB} ) \\
        &\mathrel{\phantom{=}} + \frac{3}{2} (H^{u\alpha \beta} g^{ra} H_{a \alpha \beta}) -\frac{1}{4} (g^{ur} H^{\alpha \beta \gamma} H_{\alpha \beta \gamma})\\
        &= \frac{3}{2} (\frac{1}{r^2} \gamma^{AB} H_{ruB} H_{ruA} + \frac{1}{r^4} \gamma^{AB} \gamma^{BC} H_{rBC} H_{uAB} ) + \frac{3}{2} (\frac{1}{r^2} \gamma^{AB} H_{ruB} H_{urA}) \\ 
        &\mathrel{\phantom{=}} - \frac{1}{4} (\frac{1}{r^2} \gamma^{AB} H_{uAr} H_{rAu} + \frac{1}{r^4} \gamma^{AB} \gamma^{BC} H_{uAB} H_{rAB} \\
        &\mathrel{\phantom{=}} + 2 \frac{1}{r^4} \gamma^{AB} \gamma^{BC} H_{BCr} H_{ABu} + \frac{1}{r^4} \gamma^{AB} \gamma^{BC} H_{BCu} H_{ABr} ) \\
        &= \frac{3}{2} (\frac{2}{r^2} \gamma^{AB} H_{ruB} H_{ruA} + \frac{1}{r^4} \gamma^{AB} \gamma^{BC} H_{rBC} H_{uAB} ) \\
        &\mathrel{\phantom{=}} - \frac{1}{4} (\frac{1}{r^2} \gamma^{AB} H_{uAr} H_{rAu} + \frac{1}{r^4} \gamma^{AB} \gamma^{BC} H_{uAB} H_{rAB} \\
        &\mathrel{\phantom{=}} + 2 \frac{1}{r^4} \gamma^{AB} \gamma^{BC} H_{BCr} H_{ABu})
    \end{split}
\end{equation}

Now we can insert the components in the condition above:
\begin{equation}
    \begin{split}
        | \int_{\mathcal{I}^+} r^2 (\frac{3}{2} (\frac{2}{r^2} \gamma^{AB} H_{ruB} H_{ruA} + \frac{1}{r^4} \gamma^{AB} \gamma^{BC} H_{rBC} H_{uAB} ) \\
        - \frac{1}{4} (\frac{1}{r^2} \gamma^{AB} H_{uAr} H_{rAu} + \frac{1}{r^4} \gamma^{AB} \gamma^{BC} H_{uAB} H_{rAB} \\
        + 2 \frac{1}{r^4} \gamma^{AB} \gamma^{BC} H_{BCr} H_{ABu}) \\
        - \frac{1}{2}( \frac{3}{2} (\frac{1}{r^2} \gamma^{AB} H_{ruB} H_{ruA} + \frac{1}{r^4} \gamma^{AB} \gamma^{BC} H_{rBC} H_{uAB} )) )| < \infty
    \end{split}
\end{equation}

The fall-off conditions \ref{Hfall} follow directly.

One can also impose the condition that the charge flux  at $\mathcal{I}^+$ has to be finite $| \int_{S(u)} J_{\nu\rho} dS^{\nu \rho} | =  | \int_{S(u)} (\nabla^{\mu} H_{\mu \nu \rho}) dS^{\nu \rho} |   < \infty$ , where we integrated over the two-sphere $S(u)$ at a fixed retarded time $u$ and for large $r$.

\begin{align}
    \int_{S(u)} (* J)^{\mu\nu} dS_{\mu\nu} = \int_{S(u)} (* J)^{AB} dS_{AB} = \int_{S(u)} J^{ur} r^2 d\Gamma.
\end{align}
\begin{align}
    J^{ur} = \nabla_\mu H^{\mu u r} = - \frac{\gamma^{AB}}{r^2} (D_A H_{Bru} - \frac{1}{r} H_{BAu}).
\end{align}
This calculation reaffirms the fall-off behaviour for $H_{urA}$ and $H_{uAB}$.

\section{Fall-offs of the Kalb-Ramond Fields with Duality} \label{appendixB2}
The duality relation between three-fields and scalar zero-fields in four dimensions is $H=*d\phi$. Using this and the asymptotic behaviour of the scalar field \ref{scalfall} we can derive an asymptotic behaviour for the field strengths that is consistent with the behaviour derived from other conditions.
\begin{equation}
    \begin{split}
        H_{urA} = (*d\phi)_{urA} = \sqrt{-g} \epsilon_{urAB} g^{Ba} (d\phi)_a = r^2 \sqrt{-det\gamma_{AB}} g^{AB} \partial_A \phi \\
        =\sqrt{-det\gamma_{AB}} \frac{\partial_A \phi^{(-1)} (u, \boldsymbol{\theta})}{r} \in \mathcal{O} (r^{-1}) \\
        H_{uAB} = r^2 \sqrt{-det\gamma_{AB}} ( \frac{\partial_u \phi^{(-1)} (u, \boldsymbol{\theta})}{r} + \frac{ \phi^{(-1)} (u, \boldsymbol{\theta})}{r^2} )  \in \mathcal{O} (r) \\
        H_{rAB} = r^2 \sqrt{-det\gamma_{AB}} \frac{\phi^{(-1)} (u, \boldsymbol{\theta})}{r^2} \in \mathcal{O}(r^0)
    \end{split}
\end{equation}

From the duality relation $\frac{\sqrt{-g}}{p!} \epsilon_{\mu\nu\rho\alpha} \partial^\mu B^{\nu\rho} = \partial_\alpha \phi$ , we derive the fall-off behaviour of the gauge fields (we ommitted factors of $\sqrt{-det\gamma_{AB}}$ for simplicity):
\begin{equation}
    \begin{split}
        \partial_r B_{AB}  \sim r^2 \partial_r \phi =  \phi^{(-1)} (u, \boldsymbol{\theta}) \Rightarrow B_{AB} \in \mathcal{O} (r)\\
        \partial_r B_{uA} \sim  \partial_A \phi = \frac{\partial_A \phi^{(-1)} (u, \boldsymbol{\theta})}{r} \Rightarrow B_{uA} \in \mathcal{O} (ln(r))\\
        \partial_A B_{ur} \sim  \partial_A \phi = \frac{\partial_A \phi^{(-1)} (u, \boldsymbol{\theta})}{r} \Rightarrow B_{ur} \in \mathcal{O} (r^{-1})\\
        \partial_u B_{rA} \sim  \partial_A \phi = \frac{\partial_u \phi^{(-1)} (u, \boldsymbol{\theta})}{r} \Rightarrow B_{rA} \in \mathcal{O} (r^{-1})
    \end{split}
\end{equation}

This yields exactly the asymptotic behaviour \ref{Bradfalldual}.

\printbibliography

\chapter*{Declaration}

Hiermit erkläre ich, die vorliegende Arbeit selbständig verfasst zu haben und keine anderen als die in der Arbeit angegebenen Quellen und Hilfsmittel benutzt zu haben.

Ort, Datum der Abgabe

Unterschrift

\end{document}